\documentclass[11pt,a4paper]{article}

\usepackage[T1]{fontenc}
\usepackage[utf8]{inputenc}
\usepackage{lmodern}
\usepackage{microtype}
\usepackage{amsmath,amssymb}
\usepackage{graphicx}
\usepackage{subcaption}
\usepackage{xcolor}
\usepackage{float}
\usepackage{hyperref}
\usepackage{mathrsfs}
\usepackage{cite} 
\usepackage{authblk} 

\hypersetup{
  colorlinks=true,
  linkcolor=blue,
  citecolor=blue,
  urlcolor=blue
}

\usepackage[margin=2.5cm]{geometry}

\usepackage{xcolor}
\usepackage[normalem]{ulem}

\graphicspath{{figures/}}

\begin{document}



\def\un#1{\relax\ifmmode\@@underline#1\else
        $\@@underline{\hbox{#1}}$\relax\fi}


\let\under=\unt                 
\let\ced=\ce                    
\let\du=\du                     
\let\um=\Hu                     
\let\sll=\lp                    
\let\Sll=\Lp                    
\let\slo=\os                    
\let\Slo=\Os                    
\let\tie=\ta                    
\let\br=\ub                     


\def\a{\alpha}
\def\b{\beta}
\def\c{\chi}
\def\d{\delta}
\def\e{\epsilon}
\def\f{\phi}
\def\g{\gamma}
\def\h{\eta}
\def\i{\iota}
\def\j{\psi}
\def\k{\kappa}
\def\l{\lambda}
\def\m{\mu}
\def\n{\nu}
\def\o{\omega}
\def\p{\pi}
\def\q{\theta}
\def\r{\rho}
\def\s{\sigma}
\def\t{\tau}
\def\u{\upsilon}
\def\x{\xi}
\def\z{\zeta}
\def\D{\Delta}
\def\F{\Phi}
\def\G{\Gamma}
\def\J{\Psi}
\def\L{\Lambda}
\def\O{\Omega}
\def\P{\Pi}
\def\Q{\Theta}
\def\S{\Sigma}
\def\U{\Upsilon}
\def\X{\Xi}


\def\ve{\varepsilon}
\def\vf{\varphi}
\def\vr{\varrho}
\def\vs{\varsigma}
\def\vq{\vartheta}


\def\ca{{\cal A}}
\def\cb{{\cal B}}
\def\cc{{\cal C}}
\def\cd{{\cal D}}
\def\ce{{\cal E}}
\def\cf{{\cal F}}
\def\cg{{\cal G}}
\def\ch{{\cal H}}
\def\ci{{\cal I}}
\def\cj{{\cal J}}
\def\ck{{\cal K}}
\def\cl{{\cal L}}
\def\cm{{\cal M}}
\def\cn{{\cal N}}
\def\co{{\cal O}}
\def\cp{{\cal P}}
\def\cq{{\cal Q}}
\def\car{{\cal R}}
\def\cs{{\cal S}}
\def\ct{{\cal T}}
\def\cu{{\cal U}}
\def\cv{{\cal V}}
\def\cw{{\cal W}}
\def\cx{{\cal X}}
\def\cy{{\cal Y}}
\def\cz{{\cal Z}}


\def\Sc#1{{\hbox{\sc #1}}}      
\def\Sf#1{{\hbox{\sf #1}}}      



\def\slpa{\slash{\pa}}                            
\def\slin{\SLLash{\in}}                                   
\def\bo{{\raise-.3ex\hbox{\large$\Box$}}}               
\def\cbo{\Sc [}                                         
\def\pa{\partial}                                       
\def\de{\nabla}                                         
\def\dell{\bigtriangledown}                             
\def\su{\sum}                                           
\def\pr{\prod}                                          
\def\iff{\leftrightarrow}                               
\def\conj{{\hbox{\large *}}}                            
\def\ltap{\raisebox{-.4ex}{\rlap{$\sim$}} \raisebox{.4ex}{$<$}}   
\def\gtap{\raisebox{-.4ex}{\rlap{$\sim$}} \raisebox{.4ex}{$>$}}   
\def\TH{{\raise.2ex\hbox{$\displaystyle \bigodot$}\mskip-4.7mu \llap H \;}}
\def\face{{\raise.2ex\hbox{$\displaystyle \bigodot$}\mskip-2.2mu \llap {$\ddot
        \smile$}}}                                      
\def\dg{\sp\dagger}                                     
\def\ddg{\sp\ddagger}                                   

\font\tenex=cmex10 scaled 1200


\def\sp#1{{}^{#1}}                              
\def\sb#1{{}_{#1}}                              
\def\oldsl#1{\rlap/#1}                          
\def\slash#1{\rlap{\hbox{$\mskip 1 mu /$}}#1}      
\def\Slash#1{\rlap{\hbox{$\mskip 3 mu /$}}#1}      
\def\SLash#1{\rlap{\hbox{$\mskip 4.5 mu /$}}#1}    
\def\SLLash#1{\rlap{\hbox{$\mskip 6 mu /$}}#1}      
\def\PMMM#1{\rlap{\hbox{$\mskip 2 mu | $}}#1}   %
\def\PMM#1{\rlap{\hbox{$\mskip 4 mu ~ \mid $}}#1}       %
\def\Tilde#1{\widetilde{#1}}                    
\def\Hat#1{\widehat{#1}}                        
\def\Bar#1{\overline{#1}}                       
\def\sbar#1{\stackrel{*}{\Bar{#1}}}             
\def\bra#1{\left\langle #1\right|}              
\def\ket#1{\left| #1\right\rangle}              
\def\VEV#1{\left\langle #1\right\rangle}        
\def\abs#1{\left| #1\right|}                    
\def\leftrightarrowfill{$\mathsurround=0pt \mathord\leftarrow \mkern-6mu
        \cleaders\hbox{$\mkern-2mu \mathord- \mkern-2mu$}\hfill
        \mkern-6mu \mathord\rightarrow$}
\def\dvec#1{\vbox{\ialign{##\crcr
        \leftrightarrowfill\crcr\noalign{\kern-1pt\nointerlineskip}
        $\hfil\displaystyle{#1}\hfil$\crcr}}}           
\def\dt#1{{\buildrel {\hbox{\LARGE .}} \over {#1}}}     
\def\dtt#1{{\buildrel \bullet \over {#1}}}              
\def\der#1{{\pa \over \pa {#1}}}                
\def\fder#1{{\d \over \d {#1}}}                 


\def\frac#1#2{{\textstyle{#1\over\vphantom2\smash{\raise.20ex
        \hbox{$\scriptstyle{#2}$}}}}}                   
\def\half{\frac12}                                        
\def\sfrac#1#2{{\vphantom1\smash{\lower.5ex\hbox{\small$#1$}}\over
        \vphantom1\smash{\raise.4ex\hbox{\small$#2$}}}} 
\def\bfrac#1#2{{\vphantom1\smash{\lower.5ex\hbox{$#1$}}\over
        \vphantom1\smash{\raise.3ex\hbox{$#2$}}}}       
\def\afrac#1#2{{\vphantom1\smash{\lower.5ex\hbox{$#1$}}\over#2}}    
\def\partder#1#2{{\partial #1\over\partial #2}}   
\def\parvar#1#2{{\d #1\over \d #2}}               
\def\secder#1#2#3{{\partial^2 #1\over\partial #2 \partial #3}}  
\def\on#1#2{\mathop{\null#2}\limits^{#1}}               
\def\bvec#1{\on\leftarrow{#1}}                  
\def\oover#1{\on\circ{#1}}                              

\def\[{\lfloor{\hskip 0.35pt}\!\!\!\lceil}
\def\]{\rfloor{\hskip 0.35pt}\!\!\!\rceil}
\def\Lag{{\cal L}}
\def\du#1#2{_{#1}{}^{#2}}
\def\ud#1#2{^{#1}{}_{#2}}
\def\dud#1#2#3{_{#1}{}^{#2}{}_{#3}}
\def\udu#1#2#3{^{#1}{}_{#2}{}^{#3}}
\def\calD{{\cal D}}
\def\calM{{\cal M}}

\def\szet{{${\scriptstyle \b}$}}
\def\ulA{{\un A}}
\def\ulM{{\underline M}}
\def\cdm{{\Sc D}_{--}}
\def\cdp{{\Sc D}_{++}}
\def\vTheta{\check\Theta}
\def\fracm#1#2{\hbox{\large{${\frac{{#1}}{{#2}}}$}}}
\def\ha{{\fracmm12}}
\def\tr{{\rm tr}}
\def\Tr{{\rm Tr}}
\def\itrema{$\ddot{\scriptstyle 1}$}
\def\ula{{\underline a}} \def\ulb{{\underline b}} \def\ulc{{\underline c}}
\def\uld{{\underline d}} \def\ule{{\underline e}} \def\ulf{{\underline f}}
\def\ulg{{\underline g}}
\def\items#1{\\ \item{[#1]}}
\def\ul{\underline}
\def\un{\underline}
\def\fracmm#1#2{{{#1}\over{#2}}}
\def\footnotew#1{\footnote{\hsize=6.5in {#1}}}
\def\low#1{{\raise -3pt\hbox{${\hskip 0.75pt}\!_{#1}$}}}

\def\Dot#1{\buildrel{_{_{\hskip 0.01in}\bullet}}\over{#1}}
\def\dt#1{\Dot{#1}}

\def\DDot#1{\buildrel{_{_{\hskip 0.01in}\bullet\bullet}}\over{#1}}
\def\ddt#1{\DDot{#1}}

\def\DDDot#1{\buildrel{_{_{\hskip 0.01in}\bullet\bullet\bullet}}\over{#1}}
\def\dddt#1{\DDDot{#1}}

\def\DDDDot#1{\buildrel{_{_{\hskip 
0.01in}\bullet\bullet\bullet\bullet}}\over{#1}}
\def\ddddt#1{\DDDDot{#1}}

\def\Tilde#1{{\widetilde{#1}}\hskip 0.015in}
\def\Hat#1{\widehat{#1}}


\newskip\humongous \humongous=0pt plus 1000pt minus 1000pt
\def\caja{\mathsurround=0pt}
\def\eqalign#1{\,\vcenter{\openup2\jot \caja
        \ialign{\strut \hfil$\displaystyle{##}$&$
        \displaystyle{{}##}$\hfil\crcr#1\crcr}}\,}
\newif\ifdtup
\def\panorama{\global\dtuptrue \openup2\jot \caja
        \everycr{\noalign{\ifdtup \global\dtupfalse
        \vskip-\lineskiplimit \vskip\normallineskiplimit
        \else \penalty\interdisplaylinepenalty \fi}}}
\def\li#1{\panorama \tabskip=\humongous                         
        \halign to\displaywidth{\hfil$\displaystyle{##}$
        \tabskip=0pt&$\displaystyle{{}##}$\hfil
        \tabskip=\humongous&\llap{$##$}\tabskip=0pt
        \crcr#1\crcr}}
\def\eqalignnotwo#1{\panorama \tabskip=\humongous
        \halign to\displaywidth{\hfil$\displaystyle{##}$
        \tabskip=0pt&$\displaystyle{{}##}$
        \tabskip=0pt&$\displaystyle{{}##}$\hfil
        \tabskip=\humongous&\llap{$##$}\tabskip=0pt
        \crcr#1\crcr}}


\def\eV{\,{\rm eV}}
\def\keV{\,{\rm keV}}
\def\MeV{\,{\rm MeV}}
\def\GeV{\,{\rm GeV}}
\def\TeV{\,{\rm TeV}}
\def\sv{\left<\sigma v\right>}
\def\({\left(}
\def\){\right)}
\def\cm{{\,\rm cm}}
\def\K{{\,\rm K}}
\def\kpc{{\,\rm kpc}}
\def\beq{\begin{equation}}
\def\eeq{\end{equation}}
\def\bea{\begin{eqnarray}}
\def\eea{\end{eqnarray}}


\newcommand{\be}{\begin{equation}}
\newcommand{\ee}{\end{equation}}
\newcommand{\nbe}{\begin{equation*}}
\newcommand{\nee}{\end{equation*}}
\newcommand{\overbar}[1]{\mkern1.5mu\overline{\mkern-1.5mu#1\mkern-1.5mu}\mkern 1.5mu}
\newcommand{\fr}{\frac}
\newcommand{\lb}{\label}

\thispagestyle{empty}

\hbox to\hsize{
\vbox{\noindent August 2026 \hfill IPMU26-0018 \\
\noindent  revised version  \hfill }}

\begin{center}
    \noindent {\Large \bfseries {Higgs boson mass and thermal wino dark matter \\
    from  Starobinsky supergravity with the MSSM}\par}
   
    \vspace{0.5cm}
   
    Daniel Frolovsky$^{a,b,*}$, Alexander Belyaev$^{c,d,\dagger}$, and Sergei V. Ketov$^{e,f,g,\ddagger}$
   
    \vspace{0.4cm}
   
    {\itshape \small
    $^a$ Institute for Theoretical Physics, Utrecht University,\\ Princetonplein 5, 3584 CC Utrecht, Netherlands \\[1.5mm]
    $^b$ Dutch Institute for Emergent Phenomena, Netherlands \\[1.5mm]
    $^c$ School of Physics and Astronomy, University of Southampton,\\ Highfield, Southampton SO17 1BJ, UK \\[1.5mm]
    $^d$ Particle Physics Department, Rutherford Appleton Laboratory,\\ Chilton, Didcot, Oxon OX11 0QX, UK \\[1.5mm]
    $^e$ Department of Physics, Faculty of Science, Tokyo Metropolitan University, \\ 1-1 Minami-ohsawa, Tokyo 192-0397, Japan \\[1.5mm]
    $^f$ International Institute of Physics, Natal 1613, \\ Rio Grande do Norte 59078-970, Brazil\\[1.5mm]
    $^g$ Kavli Institute for the Physics and Mathematics of the Universe (WPI),\\ The University of Tokyo Institutes for Advanced Study, Chiba 277-8583, Japan \\[1.5mm]
     \par
    }
\end{center}

\vspace{0.1cm}

{
\renewcommand{\thefootnote}{}
\footnotetext{$^\ast$ d.frolovskiy@uu.nl, the corresponding author}
\footnotetext{$^\dagger$ a.belyaev@soton.ac.uk}
\footnotetext{$^\ddagger$ ketov@tmu.ac.jp}
}
\setcounter{footnote}{0}
\renewcommand{\thefootnote}{\arabic{footnote}}

\begin{abstract}
\noindent
We propose a framework connecting cosmic microwave background (CMB) observables with high-energy particle phenomenology, based on Starobinsky supergravity coupled to the Minimal Supersymmetric Standard Model (MSSM). Cosmic inflation and supersymmetry (SUSY) breaking occur within the hidden sector emerging from Starobinsky supergravity. The inflationary scale fixes the characteristic mass scale of the hidden sector, which determines the MSSM soft terms through gravitational mediation of SUSY breaking. The same hidden sector can also dynamically generate a high-scale $\mu$ term. The resulting MSSM spectrum fixes the high-scale threshold corrections and the boundary conditions for the renormalisation-group (RG) evolution of the Higgs quartic coupling. Three-loop RG evolution of the Higgs quartic coupling gives a Higgs boson mass consistent with the measured value within the theoretical and experimental uncertainties, thereby linking the amplitude of primordial scalar perturbations to the Higgs boson mass. With conserved R-parity, the lightest supersymmetric particle is stable, and the minimal framework developed here points to a nearly pure neutral wino as the dark-matter (DM) candidate. In the thermal-relic scenario, the observed abundance selects a wino mass of about $3~{\rm TeV}$. Recent gamma-ray analyses exclude it as the sole DM component, even for conservative cored Galactic profiles. Viable alternatives include a subdominant wino fraction or late entropy dilution.  Its loop-induced spin-independent wino-nucleon scattering cross section lies below the current sensitivity of the LUX-ZEPLIN experiment, but within the projected reach of next-generation multi-ton liquid-xenon detectors. Electroweak radiative corrections generate a small mass splitting between the charged and neutral wino states, leading to a long-lived charged wino and the characteristic disappearing-track signature.  A future $100~{\rm TeV}$ proton collider can discover the disappearing-track signal from the corresponding  pure-wino state or exclude this scenario, independently of its cosmological abundance.
\end{abstract}

\clearpage
\tableofcontents
\clearpage
\section{Introduction}

One of the outstanding problems in high-energy physics is to connect the Standard Model (SM) of particle physics with cosmological inflation in the early Universe. On the one hand, the SM is the standard theoretical framework describing strong, weak and electromagnetic interactions of elementary particles, which was tested on the electroweak (EW) scale with high precision. However, as is well known, the SM is phenomenologically and theoretically incomplete because it does not include  gravity and a particle candidate for dark matter (DM), it does not explain baryon asymmetry of the Universe and non-vanishing  neutrino masses as well. The SM breaks down at very high energies when approaching the Grand Unification scale, thus requiring new physics beyond the SM. On the other hand, inflation provides a window into new physics beyond the SM, which is testable due to observations of the cosmic microwave background (CMB) radiation with high precision also. Inflation eliminates conceptual shortcomings of the standard (Einstein-Friedmann) cosmology such as the horizon and flatness problems, and provides a mechanism for generation of primordial density perturbations leading to the structure formation in the Universe. Connecting inflation to the SM is also needed for explaining the origin of elementary particles during reheating after inflation.

The existing approaches to inflation and reheating are highly model-dependent, see e.g., Refs.~\cite{Martin:2013tda,Lozanov:2019jxc,Martin:2024qnn,Kallosh:2025ijd} for a review and Refs.~\cite{Bezrukov:2007ep,DeSimone:2008ei,Bezrukov:2010jz,Hertzberg:2011rc} for the earlier attempts to connect inflation to the SM.  In this paper, we assume two main underlying concepts, namely, Starobinsky inflation \cite{Starobinsky:1980te,Ketov:2019toi,Ketov:2025nkr} and local supersymmetry \cite{Gates:1983nr,Wess:1992cp,Buchbinder:1998twe}.

Starobinsky inflation can be described in two equivalent pictures known as modified gravity and scalar-tensor gravity. These pictures are related by a Weyl transformation of spacetime metric between the so-called Jordan and Einstein frames, respectively. The modified gravity picture has the higher derivatives and the hidden scalar (inflaton) degree of freedom, whereas in the Einstein frame there are no higher derivatives and inflaton is manifest but the gravitational origin of the inflaton potential is hidden. 
The Starobinsky model based on the modified gravity action 
\be \lb{Stara}
S = \int d^4x \sqrt{-g} \left(\fracmm{M^2_{\rm P}}{2}R+\fracmm{M^2_{\rm P}}{12m_{\rm S}^2}R^2\right)
\ee
is one of the most successful models of inflation, in excellent agreement with CMB observations by the Planck mission \cite{BICEP:2021xfz}. This is more than just a good fit because the Starobinsky model is simple, geometrical, ghost-free and tightly constrained, being entirely based on gravitational interactions. It explains the origin of inflaton and the approximate flatness of its scalar potential in the Einstein frame due to scale invariance  of the dominant $R^2$ term (in the Jordan frame)  during inflation, and predicts the cosmological tilts as well. The basic Starobinsky model  can be extended by the subleading (of the higher order in the spacetime curvature) terms leading to the higher powers of $(H^2_{\rm inf}/m_{\rm S}^2)$ with the Hubble scale $H_{\rm inf}$ during inflation, which may be needed for a better agreement \cite{Addazi:2025qra} with more recent CMB observations by Atacama Cosmology Telescope (ACT) \cite{ACT:2025fju,ACT:2025tim} and South Pole Telescope (SPT) \cite{SPT-3G:2025bzu}.   The mass scale $m_{\rm inf}$ is determined by the amplitude of primordial scalar perturbations that also fix the energy scale of inflation at $\mathcal{O}(10^{13})$ GeV.

Supersymmetry (SUSY) is the fundamental symmetry relating bosons and fermions, and it is the well-known candidate for new physics beyond the SM, while local SUSY or supergravity automatically implies general relativity. In SUSY, all observed particles have superpartners (sparticles) inside irreducible SUSY multiplets. Since no sparticles were observed so far, SUSY must be spontaneously broken.  With conserved R-parity, the lightest sparticle (LSP) is a good candidate for DM. 

Supersymmetrising Starobinsky inflation implies a high SUSY scale close to the Starobinsky inflation scale, thus avoiding no-SUSY constraints from TeV physics and Big Bang Nucleosynthesis (BBN). The success of Starobinsky inflation motivates Starobinsky supergravity as the viable proposal for the gravitational effective action describing high scale inflation. A supergravity realisation of Starobinsky inflation may also determine the structure of the hidden sector together with a specific mechanism of spontaneous SUSY breaking and its mediation to the visible sector.  Most of the literature describes that in the Einstein frame, see e.g., Refs.~\cite{Ellis:2018zya,Ema:2024sit,Antoniadis:2024ypf,Ellis:2025bzi,Ellis:2026ceb}. In this paper, we first supersymmetrise the Starobinsky model in the Jordan frame by using the standard (old-minimal) set of supergravity fields \cite{Cecotti:1987sa,Gates:2009hu,Ketov:2010qz,Farakos:2013cqa,Ketov:2013dfa,Kehagias:2013mya,Addazi:2017rkc}. Due to the presence of  the higher derivatives in the Starobinsky supergravity, the "auxiliary" fields of the old-minimal off-shell supergravity multiplet become physical (propagating) degrees of freedom, which together with other hidden physical degrees of freedom form two chiral supermultiplets called the inflaton supermultiplet ${\cal T}$ and the goldstino supermultiplet ${\cal S}$ in the Einstein picture.  All these hidden degrees of freedom become manifest after a transformation of the Starobinsky supergravity from the Jordan frame to the Einstein frame, and, hence, we consider them as the supergravitational physical degrees of freedom forming the hidden sector where spontaneous SUSY breaking occurs due to the non-vanishing vacuum expectation value (VEV) of ${\cal S}$. As a result, the inflationary dynamics, the hidden sector and SUSY breaking have the single geometrical origin by emerging from the Starobinsky supergravity alone. This approach was pioneered by Hindawi, Ovrut and Waldram in 1995 \cite{Hindawi:1995qa}, and further developed by Dalianis, Farakos, Kehagias, Riotto and Unge in 2014 \cite{Dalianis:2014aya}.

Connecting inflation to the SM with gravity or to the MSSM with supergravity can be done in several different ways, either in the Einstein frame, see e.g., Refs.~\cite{Einhorn:2009bh,Hamaguchi:2014mza,Pallis:2016mvm,Dudas:2017rpa,Dudas:2017kfz,Pallis:2018xmt,Pallis:2023aom} or in the Jordan frame, see e.g., Refs.~\cite{Gorbunov:2010bn,Ferrara:2010yw,Jeong:2023zrv}, leading to different results and observable predictions. In
this paper, we supersymmetrize Starobinsky inflation in the Jordan frame, but add the MSSM in the Einstein frame, without setting  the
hidden sector "by hand". This also determines SUSY breaking scale and  soft SUSY breaking masses in the MSSM by the energy scale of inflation. The soft masses establish the common high scale SUSY breaking threshold $m_0$ acting as the upper boundary condition for the renormalisation group (RG) evolution. Below $m_0$, the effective low-energy field theory reduces to the SM. The heavy SUSY particles induce threshold corrections to the SM couplings at the matching scale. This procedure also fixes the boundary condition for the Higgs quartic coupling $\lambda(m_0)$. By employing the three-loop RG running, we relate the boundary conditions to the physical observables such as the measured Higgs boson mass. Our results become remarkably predictive and effectively eliminate free parameters typically associated with the phenomenological MSSM, including SUSY breaking scale and gravitino mass.

Another important consequence of our approach concerns the origin of the MSSM $\mu$ term. The coupling of the Higgs sector to the hidden sector of  Starobinsky supergravity naturally contains the same structural ingredient that is common to singlet-based solutions to the $\mu$ problem, where an effective $\mu$ parameter is generated when a chiral singlet superfield develops a VEV. However, in contrast to the NMSSM, we do not have to add the singlet "by hand" because it is already present as the goldstino superfield arising from the dual Einstein-frame description of Starobinsky supergravity. Its high scale VEV generates  a high scale $\mu_{\rm eff}$ of order of the gravitino mass, provided that the corresponding dimensionless coupling is suppressed. Therefore, our framework realises a high scale supergravitational version of the singlet-generated $\mu$ mechanism instead of introducing an independent low energy singlet sector.  

 In our framework, a nearly pure wino can arise as the LSP. In the standard thermal history, the relic-density calculation selects the familiar thermal-wino mass near $2.7$--$3.0~{\rm TeV}$. Realising this mass in our supergravity construction requires a high-precision cancellation between the tree-level and quantum contributions to $M_2$. Recent indirect-detection results substantially change the cosmological interpretation of this scenario: a dedicated analysis of 14 years of Fermi-LAT data excludes a pure thermal wino constituting all of the DM under the standard cosmological history, even for conservative cored Galactic profiles~\cite{Safdi:2025sfs}. The wino state can nevertheless remain phenomenologically relevant as a subdominant DM component or in a non-standard cosmological history. Its particle properties can be tested independently through direct detection and the disappearing-track signature at future colliders.
    
The paper is organised as follows. The Starobinsky supergravity setup is introduced in Sec.~2. Inflation and SUSY breaking in the
Starobinsky supergravity are considered in Sec.~3. Coupling of the Starobinsky supergravity to the MSSM is described in Sec.~4. Soft MSSM masses and RG running are derived in Sec.~5 that is the key part of this paper. Anomaly mediation of SUSY breaking and wino DM are considered in Sec.~6. Our conclusion is Sec.~7. Estimates of Higgs mass uncertainties and a comparison between \texttt{SusyHD} and \texttt{HSSUSY} codes are given in Appendix~A.

\section{Starobinsky supergravity setup}

A concise and transparent description of $N=1$ supergravity in four spacetime dimensions is provided by curved superspace \cite{Gates:1983nr,Wess:1992cp,Buchbinder:1998twe} because of its {\it manifest} local $N=1$ supersymmetry, which allows
one to construct supergravity actions without a need to check their invariance under SUSY, like in general relativity.  We restrict
ourselves to the old-minimal supergravity formulation (after the superconformal gauge fixing) that is most suitable for phenomenological applications. We use the standard (Wess-Bagger) notation of Ref.~\cite{Wess:1992cp} with the reduced Planck mass $M_{\rm P}=1$ unless stated otherwise. 

We define the Starobinsky supergravity by the following action in curved superspace of the old-minimal supergravity
(see Ref.~\cite{Ketov:2023ykf} for a review):
\begin{equation} \label{starsug}
S_{\mathcal{N}+\mathcal{F}} = \int \mathrm{d}^4x\, \mathrm{d}^4\theta\, E^{-1}\, \mathcal{N}(\mathcal{R}, \overline{\mathcal{R}}) + \left[ \int \mathrm{d}^4x\, \mathrm{d}^2\Theta 2 \mathcal{E}\,\mathcal{F}(\mathcal{R}) + \text{h.c.} \right],
\end{equation}
in terms of arbitrary non-holomorphic real potential $\mathcal{N}(\mathcal{R}, \overline{\mathcal{R}})$ and holomorphic potential $\mathcal{F}(\mathcal{R})$ depending upon the chiral superfield $\mathcal{R}$ that has the spacetime Ricci scalar $R$ amongst its field components at $\Theta^2$. The $\overline{\mathcal{R}}$ denotes the anti-chiral superfield obtained by Hermitian conjugation of $\mathcal{R}$, while $E$ and $\mathcal{E}$ are the supervielbein density and the chiral density in the full and chiral superspace, respectively.

The action (\ref{starsug}) is the manifestly supersymmetric extension of the $(R+\a R^2)$ gravity because it has only
$R$ and $R^2$ in terms of field components \cite{Ketov:2013dfa}. Equation (\ref{starsug}) defines a higher-derivative supergravity (unlike the Einstein supergravity actions in the literature) where the so-called "auxiliary" fields of the old-minimal supergravity, forming the full off-shell supergravity multiplet, become propagating in addition to graviton and gravitino. It is
customary in modified  gravity or supergravity to refer to such actions as those in the Jordan frame. The chiral $\mathcal{F}(\mathcal{R})$ supergravity \cite{Gates:2009hu,Ketov:2011rf,Ketov:2012yz,Ketov:2012se} arises when $\mathcal{N}=0$. 

The spacetime Lagrangian in the action (\ref{starsug}) can be rewritten to the chiral form as \cite{Addazi:2017rkc}
\begin{equation}
\mathcal{L} = \int \mathrm{d}^2 \Theta 2 \mathcal{E}\left[-\fracmm{1}{8}(\overline{\mathcal{D}}^2-8 \mathcal{R})\, \mathcal{N}(\mathcal{R}, \overline{\mathcal{R}})+\mathcal{F}(\mathcal{R})\right]+ \text{h.c.}~,
\end{equation}
where the $\mathcal{F}$-term can be absorbed (up to a constant) into the $\mathcal{N}$-term.  The constant can also be ignored because it leads to a spacetime cosmological constant that we ignore here. Hence, we can use
\begin{equation} \lb{DtypeL}
\mathcal{L} = \fracmm{3}{8} \int \mathrm{d}^2 \Theta 2 \mathcal{E}\left(\overline{\mathcal{D}}^2-8 \mathcal{R}\right) f(\mathcal{R}, \overline{\mathcal{R}})+\text{h.c.}
\end{equation}
that allows us to make a connection to Ref.~\cite{Dalianis:2014aya} in terms of the real $D$-type potential 
$f(\mathcal{R}, \overline{\mathcal{R}})$ alone.

Though the potential $f$ looks like a K\"ahler potential in SUSY, this is not the case here because the chiral superfield $\mathcal{R}$ is constrained in superspace supergravity. However, it can be replaced by an unconstrained chiral superfield $\mathcal{S}$ by using a Lagrange multiplier superfield $\mathcal{T}$ as
\begin{equation} \lb{chiralL}
\mathcal{L} = \left[ \fracmm{3}{8} \int \mathrm{d}^2 \Theta 2 \mathcal{E}\left(\overline{\mathcal{D}}^2-8 \mathcal{R}\right) f(\mathcal{S}, \overline{\mathcal{S}}) + 6 \int \mathrm{d}^2 \Theta 2 \mathcal{E}\, \mathcal{T}(\mathcal{S}-\mathcal{R}) \right] +\text{h.c.}
\end{equation}
On the one hand, varying with respect to  $\mathcal{T}$ yields
\begin{equation}
\mathcal{S}=\mathcal{R}
\end{equation}
that brings back the Lagrangian (\ref{DtypeL}). On the other hand, the last chiral $F$-type term in the square brackets of (\ref{chiralL}) can be rewritten to the $D$-type term in superspace, which leads to
\begin{equation}
\mathcal{L} = \fracmm{3}{8} \int \mathrm{d}^2 \Theta 2 \mathcal{E}\left(\overline{\mathcal{D}}^2-8 \mathcal{R}\right)[\mathcal{T}+\overline{\mathcal{T}}+f(\mathcal{S}, \overline{\mathcal{S}})]+ \left[ 6 \int \mathrm{d}^2 \Theta 2 \mathcal{E}\, \mathcal{T} \mathcal{S}+\text{h.c.}\right]
\end{equation}
In its turn, this Lagrangian can be rewritten to the standard form in the Einstein supergravity coupled to chiral matter,
\begin{equation} \lb{Eins}
\mathcal{L} = \fracmm{3}{8} \int \mathrm{d}^2 \Theta 2 \mathcal{E}\left(\overline{\mathcal{D}}^2-8 \mathcal{R}\right) e^{-\fracmm{1}{3} \mathcal{K}}+ \left[ \int \mathrm{d}^2 \Theta 2 \mathcal{E}\, \mathcal{W}+\text{h.c.} \right]~,
\end{equation}
with the K\"ahler potential $\mathcal{K}$ and superpotential $\mathcal{W}$ given by
\begin{equation} \lb{KandW}
\mathcal{K} = -3 \ln [\mathcal{T}+\overline{\mathcal{T}}+f(\mathcal{S}, \overline{\mathcal{S}})], \quad \mathcal{W} = 6 \mathcal{T} \mathcal{S}~~,
\end{equation}
for two chiral matter superfields $\mathcal{S}$ and $\mathcal{T}$.~\footnote{After rescaling the dimensionless superfields $\mathcal{T}$ and
$\mathcal{S}$ to the canonical dimension (one), the coupling constant  appears in front of the superpotential $\mathcal{W}$.}
The field theory (\ref{Eins}) does not have the higher derivatives and defines the equivalent (or dual) Lagrangian in the Einstein frame. Because of their supergravitational origin, the
matter superfields $\mathcal{S}$ and $\mathcal{T}$ form the hidden sector in the Einstein supergravity in addition to other
matter like the MSSM that can be added to the supergravity action (\ref{Eins}). The dependence of the K\"ahler potential
(\ref{KandW}) upon $\mathcal{T}$ has the "no-scale" form that establishes a connection to the  no-scale supergravity models \cite{Ellis:2018zya,Ema:2024sit,Antoniadis:2024ypf,Ellis:2025bzi,Ellis:2026ceb}.

The Starobinsky supergravity defines the supergravity framework in the Jordan frame, like $F(R)$ gravity
defines the framework for the Starobinsky model of inflation, see e.g., Ref.\cite{Ketov:2025cqg} for the latter.
 Describing viable Starobinsky inflation in the Starobinsky supergravity requires carefully chosen potentials $f(\mathcal{R}, \overline{\mathcal{R}})$. Unlike the modified $F(R)$ gravity that has only one (hidden) scalar physical degree of freedom, the Starobinsky supergravity has four scalar physical degrees of freedom associated  with the leading (complex) field components of two chiral matter superfields. In general, this leads to multi-field inflation in supergravity \cite{Gong:2025pbi} where all scalars contribute to the scalar potential and can easily destabilise inflation leading to a very few e-folds and large iso-curvature perturbations. Therefore, one has to stabilise three scalars out of four of them in order to get effectively single-field inflation driven by the remaining scalar called inflaton, see e.g., Ref.~\cite{Addazi:2017rkc} for details and the way to embed and stabilise the single-field Starobinsky inflation in the Starobinsky supergravity. In addition, since inflation implies positive energy, a supergravity realisation of inflation is always accompanied by spontaneous SUSY breaking that, in turn, implies the existence of a physical Nambu-Goldstone spin-1/2 fermion called goldstino. In the Starobinsky supergravity, inflaton belongs to the chiral superfield  $\mathcal{T}$, whereas
 goldstino belongs to the chiral superfield  $\mathcal{S}$. In the supergravity realisations of inflation, the goldstino superfield  
 $\mathcal{S}$ is usually stabilised at its origin by assigning its heavy mass beyond the Hubble value during inflation.

Going from the superfield formulation of a supergravity theory to its formulation in terms of the field components of the supergravity superfields is a tedious procedure requiring (i) fixing the gauge invariance against superdiffeomorphisms in curved superspace, (ii) removing auxiliary fields, and (iii) rescaling or redefining the remaining fields to get canonical normalisations of their kinetic terms \cite{Gates:1983nr,Wess:1992cp,Buchbinder:1998twe}. Fortunately, in the Einstein frame, one only needs
to know a K\"ahler potential $K$ and a superpotential $W$ or, actually, their K\"ahler-gauge-invariant combination
$G=K +\ln(W\Bar W)$.

The physical field content of the Starobinsky supergravity (in the Einstein frame, without adding matter, and after eliminating the auxiliary fields) is given by
\begin{itemize}
	\item  graviton $e^a_{\m}$ or $g_{\mu\nu}$ of spin 2 and gravitino $\psi^a_\mu$  of spin $3/2$, where both flat $(a)$ and curved $(\m)$ indices take four values,
	\item real inflaton Re$(T)$, axion (or sinflaton) Im$(T)$ of spin 0, and inflatino fermion $\chi_T$ of spin 1/2,
	\item complex sgoldstino $S$ of spin 0 and goldstino fermion $\chi_S$ of spin 1/2.
\end{itemize}
 These fields are subject to the residual gauge transformations, including local SUSY leading to equal total numbers of the bosonic and fermionic (on-shell) degrees of freedom. However, these SUSY transformations are now dependent upon the supergravity theory chosen, while their algebra is closed only on the equations of motion. 

Starobinsky inflation ending in a Minkowski vacuum can be realised  by using the function \cite{Dalianis:2014aya}
\begin{equation} \lb{f1}
f(\mathcal{R}, \overline{\mathcal{R}}) = 1 - 2\fracmm{\mathcal{R}\overline{\mathcal{R}}}{m^2} + \fracmm{1}{9}\zeta \fracmm{\mathcal{R}^2\overline{\mathcal{R}}^2}{m^4}~,
\end{equation}
where, on the right-hand-side, the first term generates the Einstein supergravity as the SUSY extension of the Einstein-Hilbert Lagrangian,  the second term generates the $R^2$-supergravity, whereas the last term with the dimensionless coupling constant $\zeta$ is needed to stabilise the additional scalar degrees of freedom (beyond inflaton) during inflation; without the quartic term, the inflationary trajectory suffers from a tachyonic instability \cite{Addazi:2017rkc}. The mass parameter $m$ in eq.~(\ref{f1}) is essentially the Starobinsky mass of order $10^{13}$ GeV.

The Starobinsky supergravity defined by Eq.~(\ref{f1}) has exact R-symmetry, and its vacuum structure is controlled by the stabilization parameter $\zeta$. To support a single-field inflation, this parameter must satisfy the stability bound 
$\zeta > 3.54$ but then inflaton ends up in a SUSY-preserving vacuum \cite{Hindawi:1995qa}. A Minkowski vacuum with broken SUSY is only possible with $\zeta=1$ but this value is incompatible with stable inflation for enough numbers of e-folds.  Hence,
in order to get a spontaneously broken SUSY after inflation in a Minkowski vacuum, one has to add R-symmetry breaking terms to the $f$-function, though without destroying Starobinsky inflation and without generating a cosmological constant. The minimal
realisation of this task amounts to adding linear terms to the $f$-function (\ref{f1}) as follows \cite{Dalianis:2014aya}:
\begin{equation}\label{fR}
f(\mathcal{R},\overline{\mathcal{R}}) = 1 + \gamma\fracmm{\mathcal{R} + \overline{\mathcal{R}}}{m} - 2\fracmm{\mathcal{R}\overline{\mathcal{R}}}{m^2} + \fracmm{1}{9}\zeta\fracmm{\mathcal{R}^2\overline{\mathcal{R}}^2}{m^4}
\end{equation}
with the new coupling constant $\gamma$ as the source of explicit R-symmetry breaking. This adjustment eliminates the constraints found in the R-symmetric case, enables the existence of a Minkowski vacuum with broken SUSY and no massless 
R-axion, while simultaneously supporting a viable single-field inflationary phase driven by the $R^2$ term like that in the Starobinsky model. In addition, this leads to heavy scalar particles in the hidden sector.


\section{Scalar potential, inflation and gravitino mass}

The scalar sector of the Lagrangian (\ref{Eins}) has the form
 \begin{equation}
	e^{-1} \mathcal{L} = -\fracmm{1}{2} R - \mathcal{K}_{i \bar{j}} \partial z^i \partial \bar{z}^j - \mathcal{V}(z^i, \bar{z}^i)~,
\end{equation}
where $\mathcal{K}_{i \bar{j}}$ is the K\"ahler metric derived from the K\"ahler potential $\mathcal{K}$, and $\mathcal{V}$ is the scalar potential of the complex scalar fields 
$z^i$ collectively representing the leading scalar field components $T$ and $S$ of the chiral superfields $\mathcal{T}$ and $\mathcal{S}$.  The scalar potential is given by the standard formula \cite{Wess:1992cp}
\begin{equation}
	\mathcal{V} = e^{\mathcal{K}} \left[ (\mathcal{K}^{-1})^{i \bar{j}} (\mathcal{D}_i \mathcal{W}) (\mathcal{D}_{\bar{j}} \overline{\mathcal{W}}) - 3 \mathcal{W} \overline{\mathcal{W}} \right]~,
\end{equation}
where $\mathcal{D}_i \mathcal{W} = \partial_i \mathcal{W} + (\partial_i \mathcal{K}) \mathcal{W}$ is the K\"ahler covariant derivative.

The choice (\ref{fR} ) for the master function $f(\mathcal{R},\overline{\mathcal{R}})$ from the preceding Sect.~2  leads to the following K\"ahler potential and superpotential:
\begin{equation} \lb{KW}
	\mathcal{K} = -3 M_{\text{P}}^2 \ln \left\{ 1 + \fracmm{\mathcal{T} + \overline{\mathcal{T}}}{M_{\text{P}}} + \gamma \fracmm{\mathcal{S} + \overline{\mathcal{S}}}{M_{\text{P}}} - 2 \fracmm{\mathcal{S} \overline{\mathcal{S}}}{M_{\text{P}}^2} + \fracmm{1}{9} \zeta \fracmm{\mathcal{S}^2 \overline{\mathcal{S}}^2}{M_{\text{P}}^4} \right\}~, \qquad  \mathcal{W} = 6m \mathcal{T} \mathcal{S}~,
\end{equation}
where we have reintroduced the reduced Planck mass $M_{\rm P}$ for more transparency from the physical viewpoint.

When parametrising the scalar fields as $T/M_{\text{P}} = t + ib$ and $S/M_{\text{P}} = s + ic$, one can demonstrate that the imaginary field components are stabilised at the origin with $\VEV{b} = \VEV{c} = 0$. Consequently, the scalar potential is reduced to a function of the real field components $t$ and $s$. We find (see also Ref.~\cite{Dalianis:2014aya})
\begin{equation} \lb{pot}
	\mathcal{V}(t,s) = 12m^2 M_{\text{P}}^2 \fracmm{s^2 \{1 - 2s^2 + \zeta s^4 - 2\gamma s\} + \fracmm{9}{2} \fracmm{[t - \gamma s - \fracmm{2}{3}s^2(3 - \zeta s^2)]^2}{(9 - 2\zeta s^2)}}{(1 - 2s^2 + \fracmm{\zeta}{9}s^4 + 2t + 2\gamma s)^2}~.
\end{equation}
Demanding this potential to have a local minimum corresponding to a Minkowski vacuum  with $\VEV{\mathcal{V}} = 0$
and spontaneously broken SUSY with $\VEV{\mathcal{D}_i \mathcal{W}}\neq 0$ constrains the parameters 
$\zeta$ and $\gamma$. Specifically, these conditions imply \cite{Dalianis:2014aya}
\begin{equation}\lb{twocon}
	\zeta = \fracmm{1 + 2s_0^2}{3s_0^4} \quad {\rm and}\quad \gamma = -2s_0 + \fracmm{2 + 4s_0^2}{3s_0}~,
\end{equation}
where $s_0 =\VEV{s}$, thus leaving only two independent parameters, namely, the VEV $s_0$ and the mass scale $m$. 

The gravitino mass evaluated at the post-inflation minimum is determined by the following supergravity formula 
\cite{Wess:1992cp}:
\begin{equation} \lb{gravitinom}
    m_{3/2}^2 = \VEV{e^G}= \VEV{e^{\mathcal{K}}\abs{\mathcal{W}}^2}= \dfrac{2187 m^2 s_0^2 (1 + 2s_0^2)^2}{8(11 - 5s_0^2)^3}~,
\end{equation}
where we have used Eqs.~(\ref{KW}), (\ref{pot}) and (\ref{twocon}). Our last equation above differs from 
the corresponding equation (4.19) in Ref.~\cite{Dalianis:2014aya}. To get a positive gravitino mass squared, the VEV of the $s$-field should obey the condition
\begin{equation} \lb{scon}
    \abs{s_0} < \sqrt{\fracmm{11}{5}}
\end{equation}
that restricts the position of the vacuum and imposes constraints on the parameters governing the potential,
\begin{equation} \lb{morecon}
\gamma > -0.54 \quad {\rm and} \quad  \zeta > 0.37~~.
\end{equation}
\begin{figure}[H] 
\centering 
\begin{minipage}{0.49\textwidth}
\centering
\includegraphics[width=\linewidth]{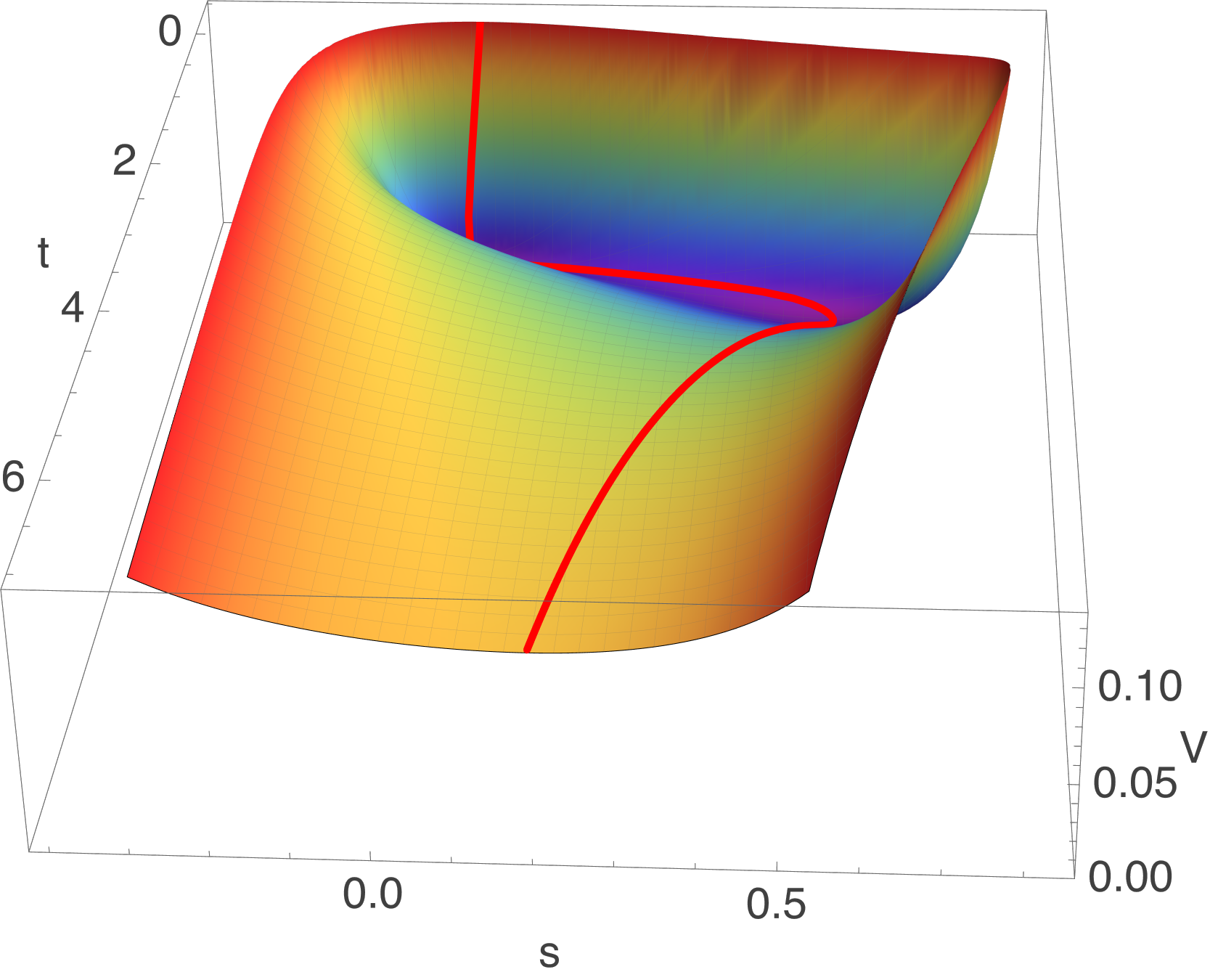}
\end{minipage}\hfill
\begin{minipage}{0.49\textwidth}
\centering
\includegraphics[width=\linewidth]{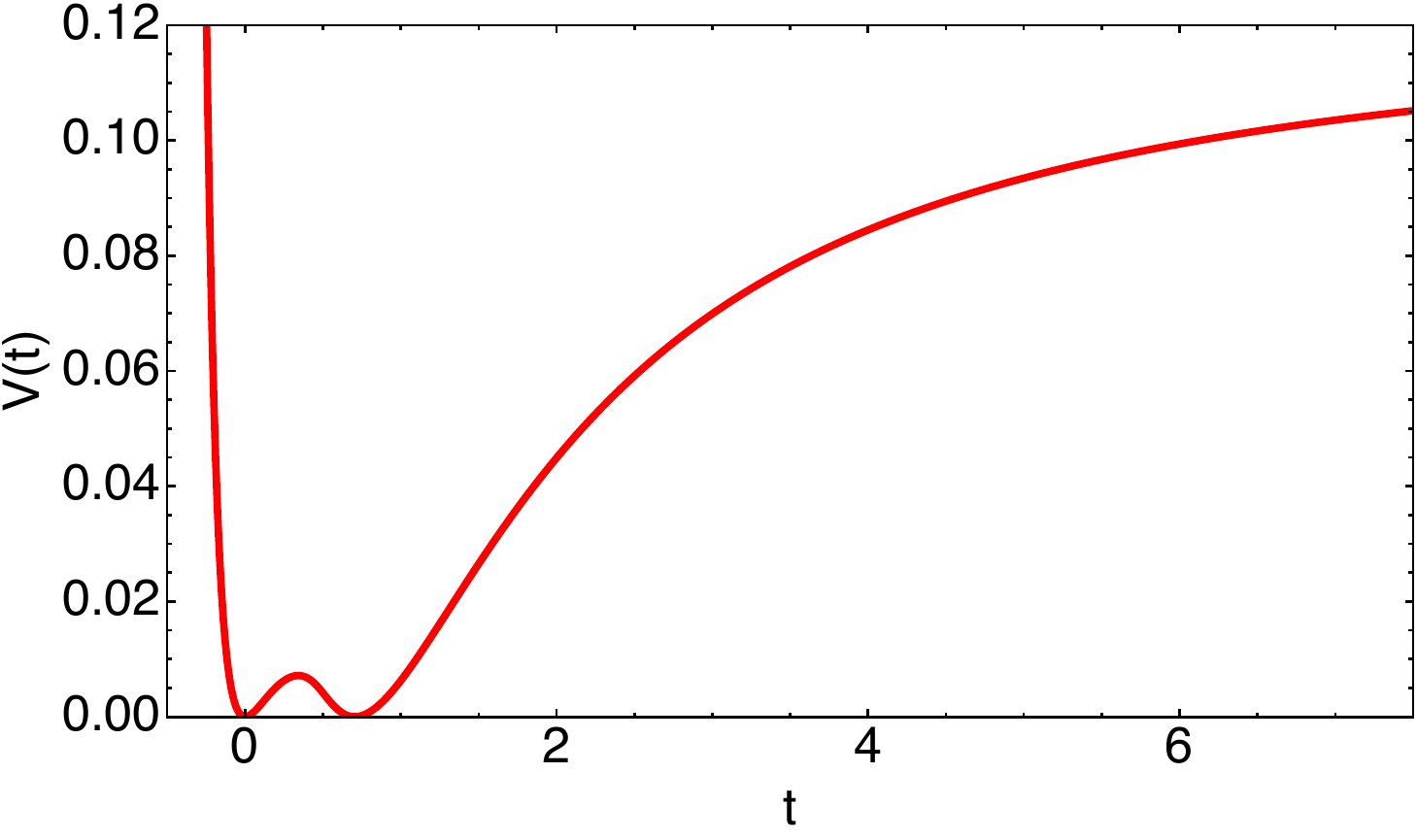}
\end{minipage}
\caption{The scalar potential $\mathcal{V}(t,s)$ overlaid by the $\partial \mathcal{V} / \partial s = 0$  trajectory is on the left-hand side. Due to its large effective mass, the field $s$ is stabilised at the minimum of the potential for any given $t$, leading to a function $s = \mathcal{Q}(t)$. A slice of the scalar potential along the $\partial \mathcal{V} / \partial s = 0$ trajectory is on the right-hand side.}
\end{figure}

During inflation, the scalar field $s$ slowly oscillates near a local minimum of the potential for any given value of the inflaton $t$.  The minimum is determined by the condition $\partial \mathcal{V}(t,s)/\partial s = 0$ that leads to $s = \mathcal{Q}(t)$ and the effective potential $\mathcal{V}(t) \equiv \mathcal{V}(t, \mathcal{Q}(t))$ leading to single-field dynamics governed by the Lagrangian
\begin{equation}\lb{quint}
e^{-1}\mathcal{L}_{\text{eff}} = -\fracmm{M_{\text{P}}^2}{2}R - \fracmm{1}{2}K(t)\partial_\mu t \partial^\mu t - \mathcal{V}(t)
\end{equation}
 Given the non-canonical kinetic term, the inflationary dynamics is described by the generalized slow-roll parameters defined by
\begin{equation}
\epsilon_K(t) = \fracmm{1}{2}M_{\text{P}}^2 \fracmm{(\mathcal{V}'(t))^2}{K(t)\mathcal{V}(t)^2} \quad {\rm and} \quad \eta_K(t) = M_{\text{P}}^2 \fracmm{\mathcal{V}''(t)}{K(t)\mathcal{V}(t)} - \fracmm{1}{2}M_{\text{P}}^2 \fracmm{K'(t)\mathcal{V}'(t)}{K(t)^2\mathcal{V}(t)}\,.
\end{equation}
The duration of inflation in terms of e-folds $N$ from a field value $t$ to the end of inflation $t_e$ defined by $\epsilon_K(t_e) = 1$ is given by
\begin{equation} \label{efolds_exact}
N(t) = \int_{t_e}^{t} \fracmm{K(\tilde{t})\mathcal{V}(\tilde{t})}{M_{\text{P}}^2 \mathcal{V}'(\tilde{t})} d\tilde{t}~.
\end{equation}
The scalar amplitude $A_s$, the scalar spectral index $n_s$, and the tensor-to-scalar ratio $r$ are evaluated at the horizon crossing on the pivot scale
denoted by $N$ as 
\begin{equation}\label{infobs}
A_s = \fracmm{\mathcal{V}(t(N))}{24 \pi^2 M_{\text{P}}^4 \epsilon_K(t(N))}\approx 2.1 \cdot 10^{-9}
\end{equation}
and
\begin{equation}
	n_s = 1 - 6\epsilon_K(t(N)) + 2\eta_K(t(N)), \quad r = 16\epsilon_K(t(N)).
\end{equation}
 
\begin{figure}[htbp]
    \centering
    \begin{subfigure}[b]{0.49\textwidth}
        \centering
        \includegraphics[width=\textwidth]{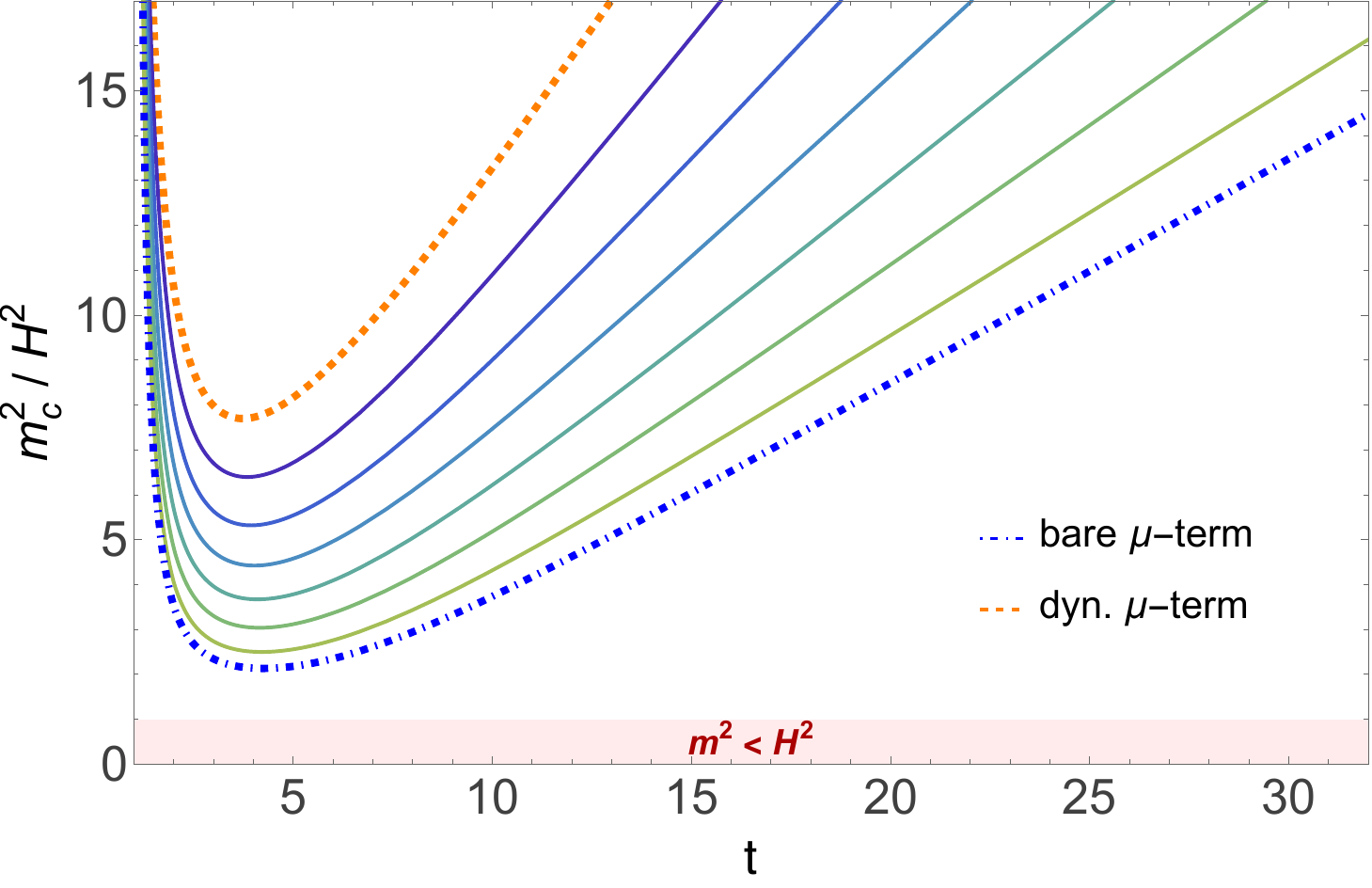}
        \caption{}
        \label{fig:mass_c}
    \end{subfigure}
    \hfill
    \begin{subfigure}[b]{0.49\textwidth}
        \centering
        \includegraphics[width=\textwidth]{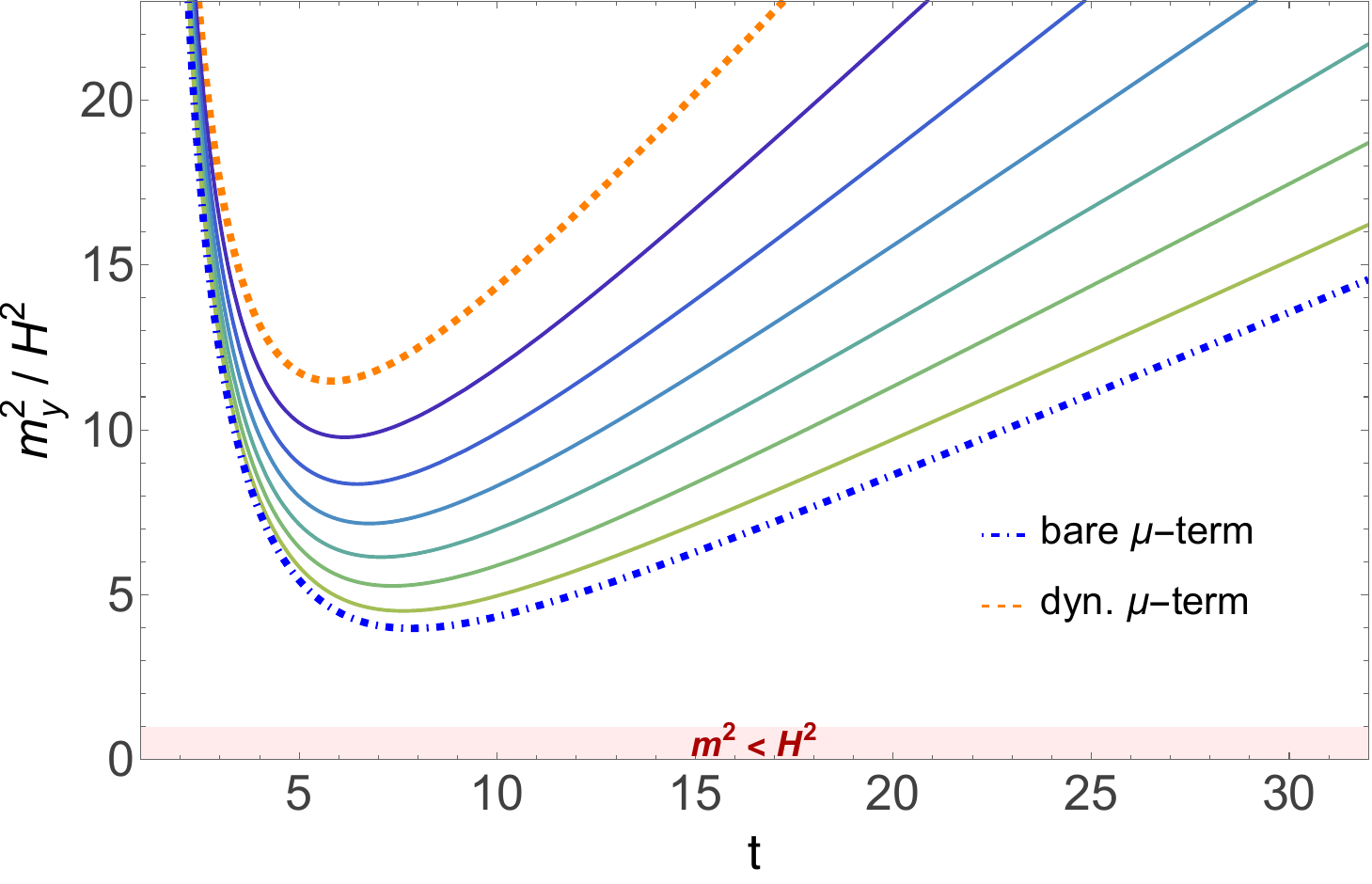}
        \caption{}
        \label{fig:mass_y}
    \end{subfigure}
    
    \vspace{0.2cm} 
    
    \begin{subfigure}[b]{0.49\textwidth}
        \centering
        \includegraphics[width=\textwidth]{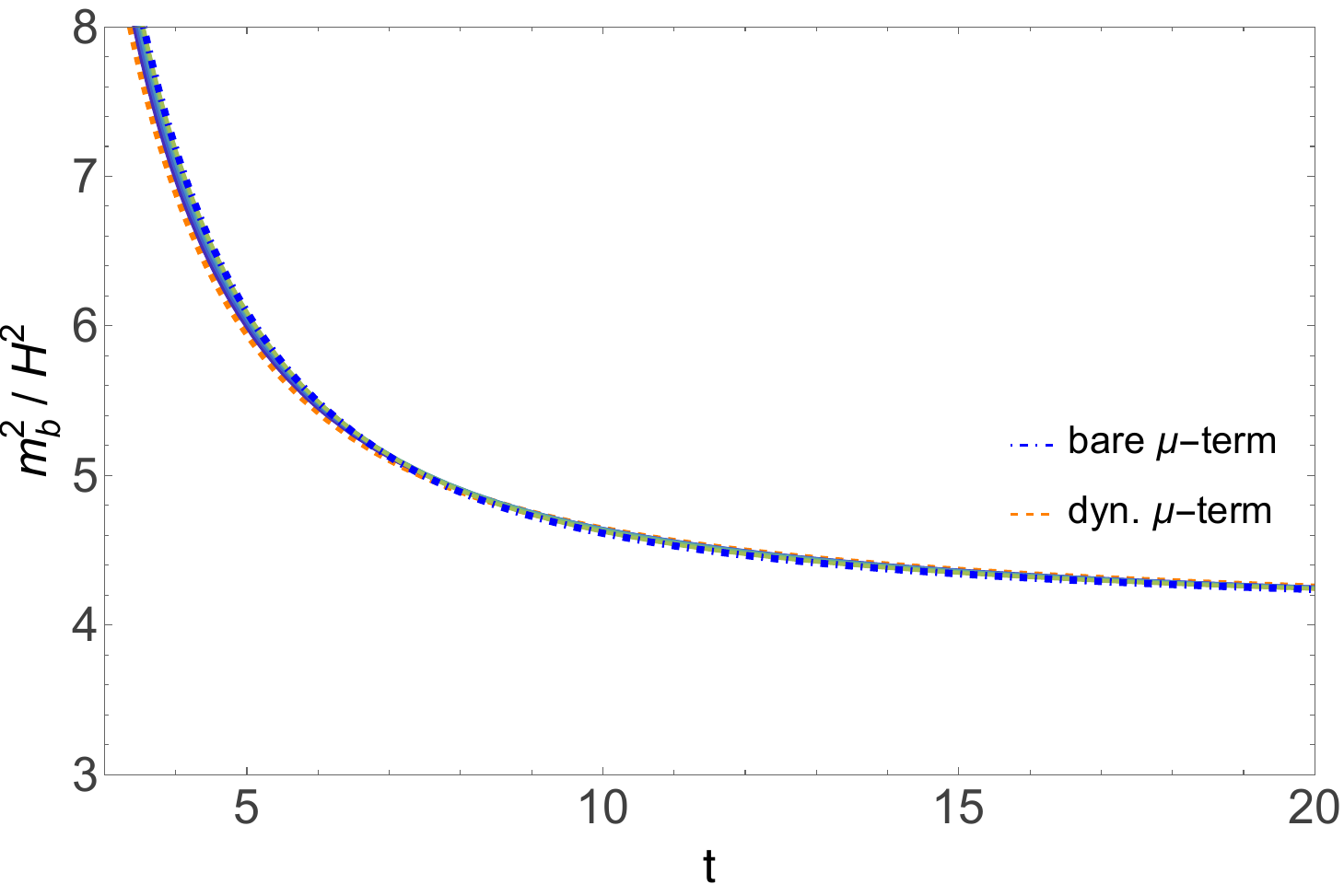}
        \caption{}
        \label{fig:mass_b}
    \end{subfigure}
    \hfill
    \begin{subfigure}[b]{0.49\textwidth}
        \centering
        \includegraphics[width=\textwidth]{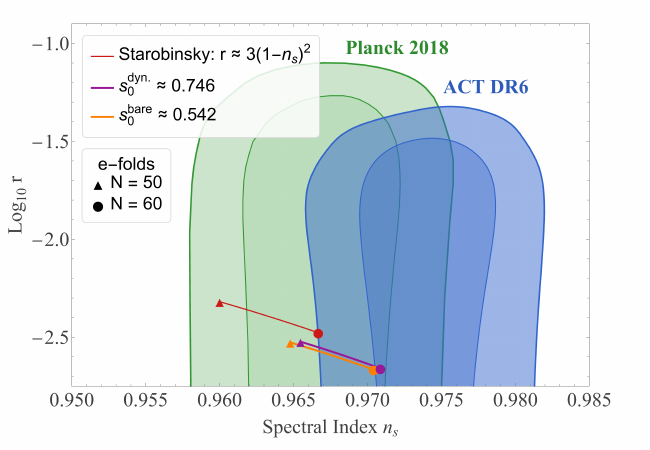}
        \caption{}
        \label{fig:ns_r_predictions}
    \end{subfigure}
    \caption{The evolution of the scalar masses in the hidden sector and the cosmological predictions.
    \textbf{(a)}--\textbf{(c)}: The effective squared masses $m_c^2/H^2$, $m_y^2/H^2$ and $m_b^2/H^2$ as the functions of the inflaton field $t$ during inflation.  The blue and red dotted lines correspond to the bare and dynamically generated $\mu$-term, respectively.   The light red shaded region at the bottom marks the boundary $m^2 < H^2$.
    \textbf{(d)}: Our predictions for the spectral index $n_s$ and the tensor-to-scalar ratio $\log_{10} r$ evaluated for e-folds $N$ in the interval $ [50, 60]$, compared against the $1\sigma$ and $2\sigma$ confidence contours from the Planck 2018 \cite{Planck:2018jri} and ACT DR6 data \cite{ACT:2025fju,ACT:2025tim}.}
    \label{fig:inflation_dynamics}
\end{figure}

To extract the physical mass in the presence of a non-canonical kinetic term for a scalar $\phi$, one should make a field redefinition to a canonical scalar field $\hat{\phi}$. In general, given a field $\phi$ with a kinetic function $\mathcal{B}(\phi)$, the transform yields the canonical mass squared $m_{\hat{\phi}}^2 = \fracmm{\langle \mathcal{V}''(\phi) \rangle}{\langle \mathcal{B} \rangle}$ to be evaluated at the potential minimum. This way we get the canonical squared masses $m_c^2/H^2$, $m_y^2/H^2$ and $m_b^2/H^2$ with $y = s - \mathcal{Q}(t)$, while the potential has a minimum at $y = 0$. The results are given in  Fig.~\ref{fig:inflation_dynamics} (a)-(c). All those canonical masses are above the Hubble scale $H_{\rm inf}$ during inflation for all relevant values of $s_0$. This confirms that the fields $b$, $c$ and $y$ are effectively frozen during inflation driven by the inflaton $t$. 

The results of our numerical calculations for the e-folds $N$ belonging to the interval $[50, 60]$ against the Planck 2018 and ACT DR6 confidence contours, and against the Starobinsky model predictions are given in Fig.~\ref{fig:inflation_dynamics} (d), being in excellent agreement with observations. The predicted values of $n_s$ slightly increase in our supergravity model against those in the original Starobinsky model toward better agreement with ACT observations at $2\sigma$ confidence level. Equation~(\ref{infobs}) determines the mass parameter $m$ from the observed CMB amplitude $A_s$ and the number of e-folds $N$. 

 The mass $m$ depends not only upon $N$ but slightly upon the goldstino VEV $s_0$ also. The value of $m$ is lower than the standard Starobinsky mass given by
\begin{equation} \label{starm}
m_{\rm S} = \sqrt{ 24\pi^2A_s} \left( \fracmm{M_{\rm P}}{N}\right)\approx
3.4 \left( \fracmm{50}{N}\right) \cdot 10^{13}~{\rm GeV}~. 
\end{equation}
The value of $m$ in our framework links the cosmological evolution to the scale of SUSY breaking in the hidden sector, where the goldstino VEV $s_0$ determines the gravitino mass $m_{3/2}$ according to Eq.~(\ref{gravitinom}).

\begin{figure}[t!]
    \centering
    \begin{subfigure}[b]{0.49\textwidth}
        \centering
        \includegraphics[width=\textwidth]{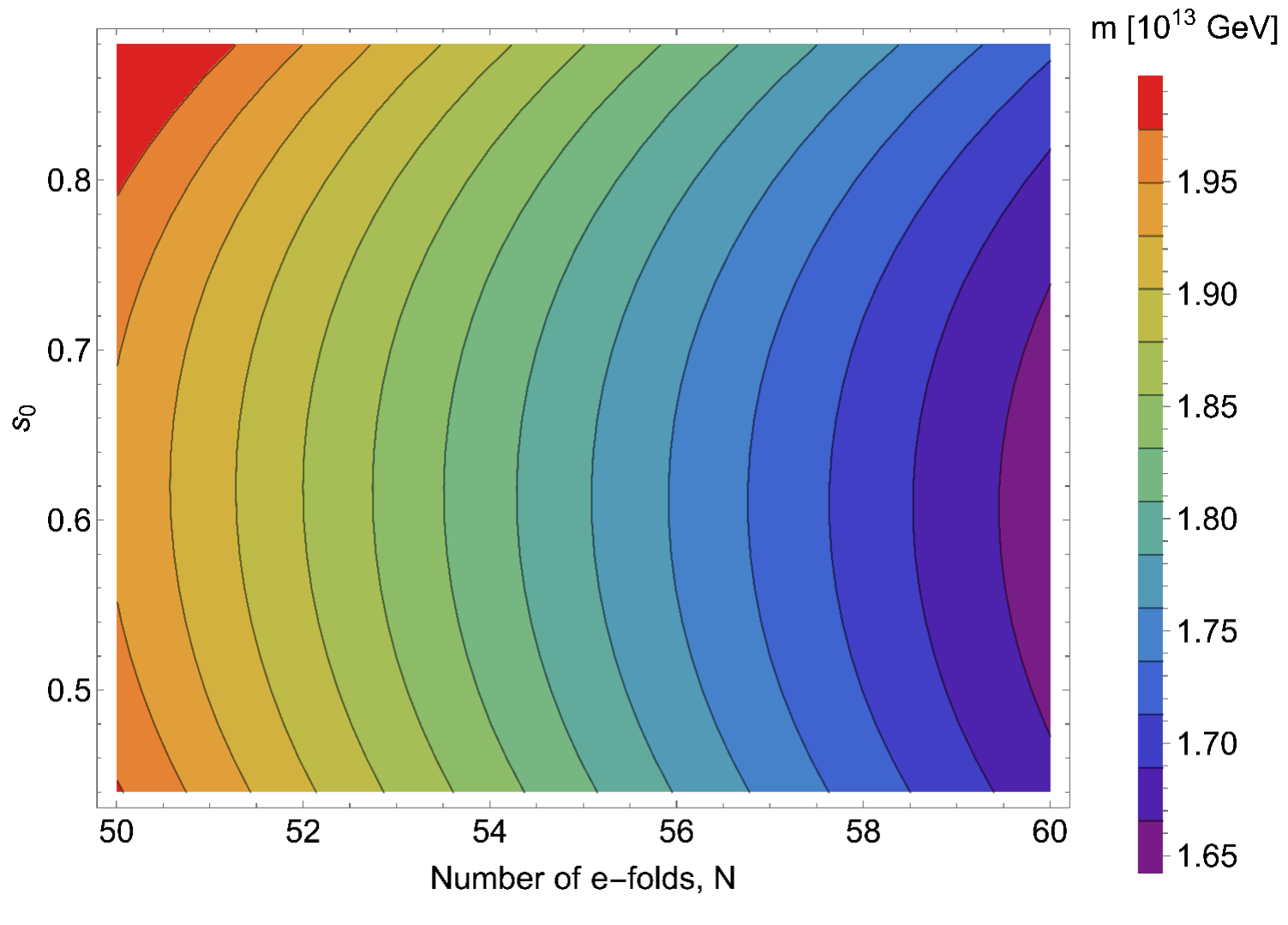}
        \caption{}
        \label{fig:inflaton_mass}
    \end{subfigure}
    \hfill
    \begin{subfigure}[b]{0.49\textwidth}
        \centering
        \includegraphics[width=\textwidth]{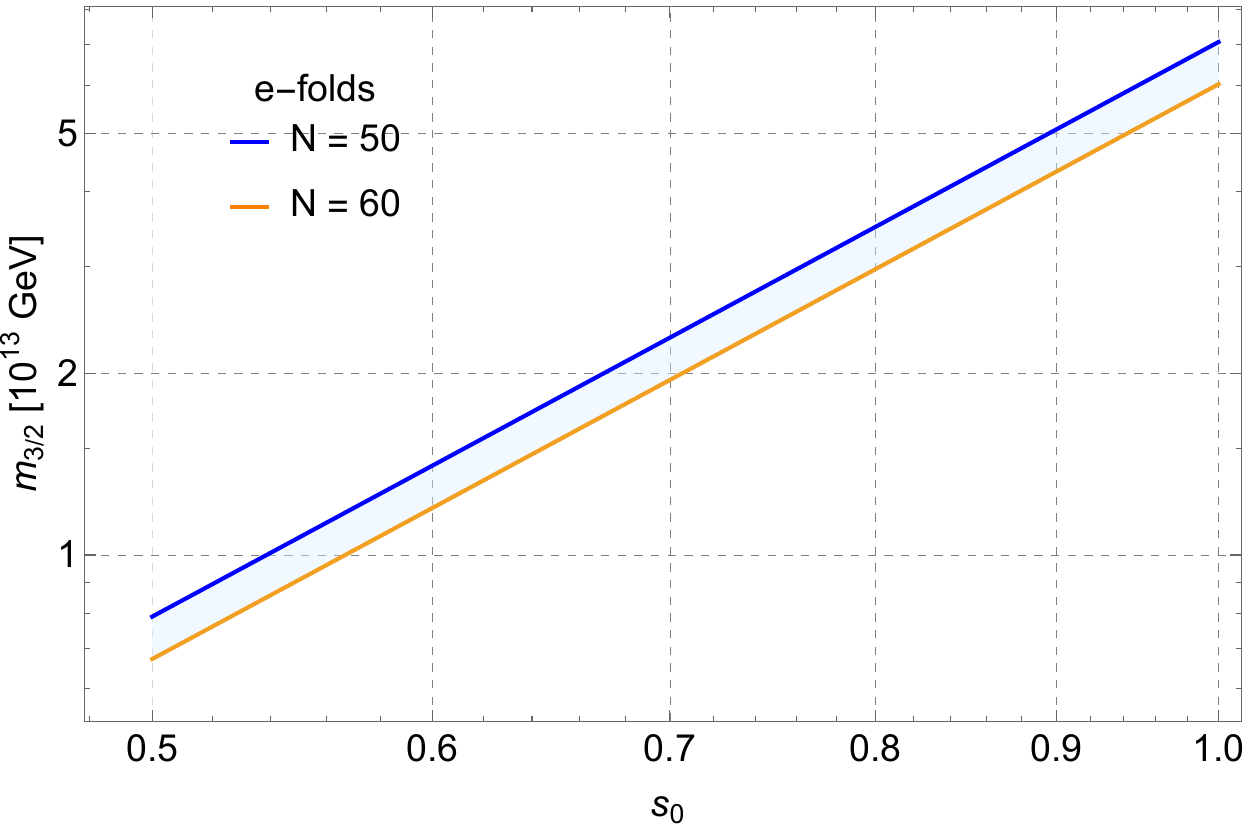}
        \caption{}
        \label{fig:gravitino_mass}
    \end{subfigure}
    \caption{
    \textbf{(a)} Numerical evaluation of the inflaton mass parameter $m$ (in units of $10^{13}$~GeV) as a function of the pivot e-folds $N$ over the interval $[50, 60]$  and the goldstino VEV $s_0$ over the scan interval $(0.438,0.886)$.
    \textbf{(b)} Gravitino mass $m_{3/2}$ (in units of $10^{13}$~GeV) as a function of the goldstino VEV $s_0$ in the $\log\log$ scale. The thin band in light blue highlights the  variation of the results across the range of e-folds in the interval $[50, 60]$.}
    \label{fig:masses_plots}
\end{figure}

Theoretical and phenomenological consistency of our approach imposes bounds on the goldstino VEV $s_0$. To have a positive gravitino mass squared and vacuum stability, $s_0$ should satisfy Eq.~(\ref{scon}) that gives the upper bound $|s_0| < \sqrt{11/5} \approx 1.483$.  A sufficient condition for kinematically closing the direct inflaton decay channel into a gravitino pair is $m_{\rm inf}<2m_{3/2}$, which gives $s_0>0.668$.

Possible small variations in the predictions for the cosmological tilts $(n_s,r)$ may also be explained within Starobinsky supergravity by adding higher powers of ${\cal R}$ and $\bar{\cal R}$, or an additional dependence upon other superfields of the superspace supergravity (perhaps involving their covariant derivatives), to the master function $f$ in Eq.~(\ref{DtypeL}); see, e.g., Refs.~\cite{Gates:2009hu,Toyama:2024ugg,Addazi:2025qra}. Yet another possible option could be non-instantaneous reheating, which increases the value of $N$ \cite{Zharov:2025zjg}.  

As is clear from Fig. \ref{fig:inflation_dynamics} (d), our predictions for $n_s$ and $r$ are in good agreement with observations. For details on the differences between the Planck and ACT constraints, see Refs. \cite{Ferreira:2025lrd, Frolovsky:2025iao}.



\section{Coupling of Starobinsky supergravity to MSSM}

There are several different ways to couple the MSSM to the Starobinsky supergravity, either in the Jordan frame or in the Einstein frame, either in terms of superfields or in terms of fields. For instance, when the superfield MSSM is coupled to the Starobinsky supergravity in the Jordan frame, this leads to the K\"ahler potential
\begin{equation}\label{jmssm}
	\mathcal{K} = -3 M_{\text{P}}^2 \ln \left( 1 + \fracmm{\mathcal{T} + \overline{\mathcal{T}}}{M_{\text{P}}} + \gamma \fracmm{\mathcal{S} + \overline{\mathcal{S}}}{M_{\text{P}}} - 2 \fracmm{\mathcal{S} \overline{\mathcal{S}}}{M_{\text{P}}^2} + \fracmm{1}{9} \zeta \fracmm{\mathcal{S}^2 \overline{\mathcal{S}}^2}{M_{\text{P}}^4} + \sum_A \fracmm{\overline{\Phi}_A \Phi_A}{M_{\text{P}}^2} \right)
\end{equation}
with non-canonical kinetic terms for all MSSM chiral superfields $\Phi_A$.

Another option is to choose the manifestly supersymmetric coupling of the MSSM matter in the \textit{Einstein} frame, in order to have the \textit{canonical} kinetic terms for the MSSM matter with a ``renormalizable'' superpotential up to third order in the chiral superfields $\Phi_A$. Then the K\"ahler potential and the superpotential are given by
\begin{equation} \label{K2}
	\mathcal{K} = -3 M_{\text{P}}^2 \ln \left( 1 + \fracmm{\mathcal{T} + \overline{\mathcal{T}}}{M_{\text{P}}} + \gamma \fracmm{\mathcal{S} + \overline{\mathcal{S}}}{M_{\text{P}}} - 2 \fracmm{\mathcal{S} \overline{\mathcal{S}}}{M_{\text{P}}^2} + \fracmm{1}{9} \zeta \fracmm{\mathcal{S}^2 \overline{\mathcal{S}}^2}{M_{\text{P}}^4} \right) + \sum_A \overline{\Phi}_A \Phi_A ,
\end{equation}
and
\begin{equation} \label{W2}
	\mathcal{W} = 6m \mathcal{T} \mathcal{S} + \mathcal{W}_{\text{MSSM}}(\Phi_A)~,
\end{equation}
where
\begin{equation} \label{mssmW}
	\mathcal{W}_{\text{MSSM}}(\Phi_A) = \mu H_u H_d + \sum_{A,B,C} y_{ABC} \Phi_A \Phi_B \Phi_C~,
\end{equation}
and the sums go over the MSSM chiral matter superfields
\begin{equation}\lb{mssmA}
	\Phi_A = (Q,L,d^c,u^c,e^c,H_d,H_u)~.
\end{equation}
The first term in Eq.~(\ref{mssmW}) is called the $\mu$-term, and $y_{ABC}$ are the MSSM Yukawa coupling constants.\footnote{Indices for three particle generations are suppressed. The non-vanishing Yukawa couplings are for $(Q,H_d,d^c)$, $(Q,H_u,u^c)$ and $(L,H_d,e^c)$. The Yukawa couplings are not derivable in supergravity but may be selected from string theory compactification \cite{Candelas:1987rx,Font:1988mm,Cheshel:2002wj}.}
 In what follows, $\mu$ should be understood either as an effective parameter in the MSSM superpotential at the matching scale or as a quantity dynamically generated by the hidden sector, as discussed below.
The chiral matter sector in the MSSM is coupled to the gauge matter sector described by vector gauge superfields, see e.g., Ref.~\cite{Aitchison:2005cf} for details.

SUSY breaking in the hidden sector can be mediated to the MSSM sector at the tree level via gravitational couplings in the field Lagrangian derived from the $\mathcal{K}$ and the $\mathcal{W}$. In turn, this generates soft SUSY breaking terms in the MSSM after decoupling of gravity at smaller scales. Equations (\ref{jmssm}) and (\ref{K2}) lead to different soft terms and different physics. In what follows, we proceed with Eq.~(\ref{K2}).

 The $\mu$-term in Eq.~(\ref{mssmW}) can be dynamically generated from interactions between the Higgs sector and the hidden sector of supergravity. This mechanism is structurally the same as singlet-based solutions of the $\mu$-problem, such as the Next-to-Minimal Supersymmetric Standard Model (NMSSM), where an effective supersymmetric Higgs mass is generated when a gauge-singlet chiral superfield develops a VEV, $\mu_{\rm eff}=\lambda \langle N\rangle$~\cite{Kim:1983dt,Ellwanger:2009dp}. The important difference is that in the present construction the singlet is not added as an extra low-energy sector. The required chiral superfield is already present as the goldstino multiplet $\mathcal{S}$ emerging from Starobinsky supergravity in the Einstein frame.

For instance, the $\mu$-term of order the gravitino mass,
$\mu \sim \mathcal{O}(m_{3/2})$, can arise from the K\"ahler potential via the standard
Giudice-Masiero mechanism~\cite{Giudice:1988yz,Casas:1992mk}. Alternatively, in our
framework the $\mu$-term can be generated from the superpotential interaction
\begin{equation} \label{gmterm}
	\fracmm{\tilde{\lambda}}{M_{\text{P}}^{n-1}} \mathcal{S}^n H_u H_d~,
\end{equation}
 where $\tilde{\lambda}$ is a dimensionless coupling. Since the scalar component of the
goldstino superfield $\mathcal{S}$ acquires the Planck-scale VEV
$\VEV{\mathcal{S}} = s_0 M_{\text{P}}$ after inflation, this interaction generates
\begin{equation}
	\mu = \tilde{\lambda} s_0^n M_{\text{P}}~.
\end{equation}
 Thus the model contains the same basic ingredient as singlet-generated $\mu$-mechanisms,
but embeds it into the supergravitational hidden sector rather than into an extra low-energy
singlet sector. Since $\VEV{\mathcal{S}}$ is of order $M_{\text{P}}$, obtaining
$\mu \sim m_{3/2} \sim 10^{13}$~GeV requires a suppressed dimensionless coupling
$\tilde{\lambda}\sim m_{3/2}/M_{\text{P}}$, up to powers of the dimensionless $s_0$.
Such a suppression can be dynamically realised if $\tilde{\lambda}$ is generated by
non-perturbative effects, for example, by instantons~\cite{Candelas:1987rx,Font:1988mm,Cheshel:2002wj}.
The phenomenologically viable value of $n$ is fixed below by the Higgs sector boundary
conditions. See Ref.~\cite{Bae:2019dgg} for a review of the $\mu$-problem and its possible
solutions.

 A direct phenomenological consequence is that the higgsinos are heavy in the high-scale solutions considered here. Since $\mu_{\rm eff}$ is tied to the gravitino mass scale, the higgsino states decouple together with the scalar superpartners and cannot provide a TeV-scale higgsino LSP.


\section{Soft MSSM parameters and RG running}

Having obtained the K\"ahler potential $\mathcal{K}$ and the superpotential $\mathcal{W}$ as sums of contributions from the hidden and MSSM sectors, one can derive soft SUSY breaking terms. To obtain the low-energy effective Lagrangian, we apply the standard procedure of taking the ``flat limit''~\cite{Brignole:1997wnc,Brignole:2010sax} by sending $M_{\rm P} \to \infty$ while keeping $m_{3/2}$ fixed. In this limit, all soft scalar masses, trilinear and bilinear couplings are derived from an expansion of the scalar potential in supergravity. When the auxiliary fields $F^m$ of the hidden sector superfields acquire non-zero VEVs, the effective mass terms for squarks, sleptons and Higgs bosons, as well as the trilinear $A$-terms and bilinear $B$-terms, are generated. The gaugino masses are generated from gaugino couplings to the $F$-terms via a holomorphic gauge kinetic function $f_a$. Communication of SUSY breaking from the hidden sector to the MSSM sector is therefore mediated by gravity at the classical level, with additional anomaly-mediated contributions generated at the quantum level.

The general setup of Refs.~\cite{Brignole:1997wnc,Brignole:2010sax} in terms of physical fields of the MSSM matter superfields is greatly simplified in our case with the K\"ahler potential in Eq.~(\ref{K2}) and the superpotential in Eq.~(\ref{W2}). In particular, the soft scalar masses are given by
\begin{equation}
m_{\bar{\alpha} \beta}^{2} = m_{3/2}^2 \delta_{\bar{\alpha} \beta} \quad {\rm or} \quad  m_0=m_{3/2}~. 
\end{equation}
The universal trilinear parameter $A$ is given by a contraction of the hidden sector F-terms with the derivatives of the K\"ahler potential, $A = F^m \partial_m \mathcal{K}$. With the K\"ahler potential in Eq.~(\ref{K2}), we find
\begin{equation}
A = m_{3/2} \left[ 3 - \fracmm{2(11 - 5s_0^2)}{3(1 + 2s_0^2)} \right] = m_{3/2} \fracmm{28s_0^2 - 13}{3(1 + 2s_0^2)}~~.
\end{equation}

In the absence of bilinear mixing terms in the K\"ahler potential, the bilinear soft parameter $B$ reduces to $B = F^m \partial_m \mathcal{K} - m_{3/2}$ and reads
\begin{equation} \lb{BA}
B = A - m_{3/2} = m_{3/2} \fracmm{2(11s_0^2 - 8)}{3(1 + 2s_0^2)}~.
\end{equation}
The physical Higgs mixing parameter $\mu_{\text{eff}}$ is derived from the superpotential coupling constant $\mu$ after rescaling as
\begin{equation} \lb{mueff_gen}
\mu_{\text{eff}} = e^{\langle \mathcal{K} \rangle / 2} \mu.
\end{equation}

The gaugino masses $M_a$ are determined by the diagonal elements of the gauge kinetic function  $f_{a}$ as~\cite{Brignole:1997wnc,Brignole:2010sax}
\begin{equation}\label{gmass}
M_a = \frac{1}{2} (\text{Re} f_a)^{-1} F^m \partial_m f_a.
\end{equation}
When adopting a linear {\it Ansatz}, $f_{a} =  (1 + c_T T + c_S S)$ with the parameters $c_T$ and $c_S$, on the tree level we get
\begin{equation} \lb{gkf}
M_a = m_{3/2} \fracmm{2(11 - 5s_0^2) \left[ 2c_T (1 + 2s_0^2) + 9c_S s_0 \right]}{9(1 + 2s_0^2) \left[ 9 + 4c_T(1 + 2s_0^2) + 9c_S s_0 \right]} =m_{1/2}~.
\end{equation}
Therefore, as may have been expected, the entire MSSM soft spectrum on the high scale is proportional to the gravitino mass $m_{3/2}$. The VEV $s_0$ is dynamically generated by the fundamental R-symmetry breaking parameter $\gamma$, while the gravitino mass $m_{3/2}$ is tied to the Starobinsky inflation scale $m \sim 10^{-5} M_P$ via the vacuum relation (\ref{gravitinom}).

To connect the high-scale SUSY predictions to the EW-scale observables,  we use two independent effective field theory (EFT) implementations, \texttt{SusyHD}~\cite{PardoVega:2015eno} and \texttt{HSSUSY} within \texttt{FlexibleSUSY}~\cite{Athron:2016fuq,Athron:2017fvs,Allanach:2018fif}. Both codes match the MSSM onto the SM at the high SUSY scale and re-sum the large logarithms generated by the wide separation between the EW and SUSY breaking scales through RG evolution. We  use \texttt{SusyHD} for the main parameter scans and \texttt{HSSUSY} as an independent cross-check of selected benchmark points and uncertainty estimates. This comparison tests the stability of the Higgs mass prediction against differences in the implementation of threshold corrections, the extraction of the low-energy SM parameters, and the treatment of the renormalisation scales. The soft spectrum above serves as the initial condition for RG running. When following the standard high-scale SUSY scenario~\cite{Bagnaschi:2014rsa}, the MSSM is directly mapped onto the SM at the scale $m_0 = m_{3/2}$. Below the matching scale, the heavy sparticles are integrated out. The boundary condition for the SM Higgs quartic coupling $\lambda(m_0)$ is determined by the MSSM tree-level relation combined with the high-scale threshold corrections $\Delta\lambda$ as
\begin{equation}\label{lam}
	\lambda(m_0)=\fracmm{1}{4}\left[g_2^2(m_0)+\fracmm{3}{5} g_1^2(m_0)\right] \cos ^2 2 \beta +\Delta\lambda~.
\end{equation}
The threshold corrections are sensitive to the stop-mixing parameter $X_t = A_t - \mu_{\text{eff}} \cot \beta$ and include the full one-loop contributions and the leading two-loop supersymmetric contributions. The stop-mixing parameter $X_t$ must be constrained in order to prevent the formation of colour-and-charge-breaking (CCB) minima along the squark field directions~\cite{Bagnaschi:2014rsa}. More specifically, demanding that a CCB minimum is not deeper than the standard EW minimum dictates an upper bound on the stop mixing. Since all supersymmetric scalars in our framework acquire the universal high-scale mass $m_{3/2}$, the stability condition takes the form
\begin{equation}
\fracmm{(A_t - \mu_{\text{eff}} \cot \beta)^2}{m_{3/2}^2} < 8 - \fracmm{2}{\sin^2 \beta}~.
\end{equation}

The Higgs coupling $\lambda$ is evolved down to the EW scale according to the three-loop RG equations. This procedure effectively resums large logarithms of the form $\ln(m_{3/2}/m_t)$. The physical Higgs boson pole mass $M_h$ is determined at the top mass scale by the relation
\begin{equation}\label{hmass}
M_h^2 = v^2 \left[ \lambda(m_t) + \delta\lambda(m_t) \right] ,
\end{equation}
where $v \simeq 246.22$~GeV is the Higgs VEV. The term $\delta\lambda(m_t)$ represents the finite threshold corrections computed up to two loops, which are required for matching the running parameters with the physical mass. The SM Higgs doublet is given by a linear combination of the two MSSM Higgs doublets $H_u$ and $H_d$ as
\begin{equation} \lb{Hmix}
H_{\rm SM}= H_u\sin\b + H_d^{\dg}\cos\b 
\end{equation}
with the mixing angle $\b$. 

In our framework, both Higgs doublets acquire the large soft masses $m_{H_u}=m_{H_d}=m_{3/2}$. For a light SM Higgs doublet to exist well below the SUSY breaking scale, the tree-level Higgs mass-squared matrix should have one eigenvalue much smaller than the characteristic SUSY scale, $m_{\rm light}^2 \ll m_{3/2}^2$, whereas the other eigenvalue stays heavy, $m_{\rm heavy}^2 \sim m_{3/2}^2$. The determinant of the Higgs sector mass-squared matrix can be expressed in terms of these eigenvalues as follows:
\begin{equation}\lb{detcon}
\det(\mathcal{M}_H^2) = (m_{H_u}^2 + \mu_{\text{eff}}^2)(m_{H_d}^2 + \mu_{\text{eff}}^2) - (B\mu_{\text{eff}})^2 = m_{\text{light}}^2 m_{\text{heavy}}^2~.
\end{equation}
Thus, the existence of a light SM-like Higgs doublet requires a cancellation between the diagonal and off-diagonal contributions to $\mathcal{M}_H^2$, so that $\det(\mathcal{M}_H^2)\ll m_{3/2}^4$. This is the usual EW fine-tuning in high-scale SUSY. At leading order, the cancellation can be formulated as the high-scale boundary condition
\begin{equation}\lb{detcon0}
\det(\mathcal{M}_H^2)=0
\end{equation}
with high precision. As shown below, in our framework this leads to an equation on the goldstino VEV $s_0$. 
After substituting the universal soft scalar masses $m_{H_u} = m_{H_d} = m_{3/2}$ into Eq.~(\ref{detcon0}), we get
\begin{equation}\label{detcon1}
	m_{3/2}^2 + \mu_{\text{eff}}^2 = \abs{B\mu_{\text{eff}}}\,.
\end{equation}
By dividing both sides by $m_{3/2}\abs{\mu_{\text{eff}}}$, this relation can be rewritten in terms of the dimensionless ratio $x \equiv \abs{\mu_{\text{eff}}}/m_{3/2} > 0$ as
\begin{equation}
	x + \fracmm{1}{x} = \fracmm{\abs{B}}{m_{3/2}}~.
\end{equation}
Since $x + 1/x \ge 2$ for any $x > 0$, this imposes a lower bound on the bilinear soft parameter,
\begin{equation}
	\abs{B} \ge 2m_{3/2}\,.
\end{equation}
Inserting Eq.~(\ref{detcon1}) into the minimization condition for the Higgs potential,
\begin{equation} \label{minHcon}
\sin 2\beta = \fracmm{2B\mu_{\text{eff}}}{m_{H_u}^2 + m_{H_d}^2 + 2\mu_{\text{eff}}^2}~,
\end{equation}
implies $\abs{\sin 2\beta} = 1$ or, equivalently, $\abs{\tan \beta} = 1$ up to small corrections of order $\mathcal{O}(m_{\text{light}}/m_{3/2})$. Therefore, as implied by Eq.~(\ref{lam}) and Eq.~(\ref{hmass}), the physical Higgs boson mass is almost entirely generated by radiative threshold corrections and the subsequent RG running.

Regarding the high-scale boundary condition for the stop-mixing parameter $X_t = A_t - \mu_{\text{eff}} \cot \beta$, we adopt the minimal choice $X_t = 0$. This condition is imposed to suppress the high-scale threshold corrections to the Higgs quartic coupling. Assuming $X_t = 0$ also ensures the stability of the  vacuum along the squark directions and avoids CCB minima~\cite{Bagnaschi:2014rsa}. Given the universality ($A_t = A$), this assumption yields $\mu_{\text{eff}} = A \tan \beta$. Inserting this relation back into Eq.~(\ref{minHcon}), we can safely remove the absolute value in Eq.~(\ref{detcon1}), yielding $BA > 0$ and $\tan \beta = 1$.

Finally, given $\tan \beta = 1$, the condition $X_t = 0$ dictates $A = \mu_{\text{eff}}$. Equations (\ref{BA}) and (\ref{detcon1}) then imply 
\begin{equation} \lb{Asol}
m_{3/2}^2 + A^2 = A(A-m_{3/2}) \quad {\rm and,~hence,} \quad A = -m_{3/2}~,
\end{equation}
which fixes the parameter $\mu_{\text{eff}} = -m_{3/2}$. Then taking into account Eq.~(\ref{mueff_gen}) allows us to express the parameter $\mu$ in Eq.~(\ref{mssmW}) in terms of the mass scale $m$ and the VEV of $\mathcal{S}$ as
\begin{equation} \lb{muend}
	\mu = -\fracmm{8}{3} m s_0 (1 + 2s_0^2)~.
\end{equation}

The relation $A = -m_{3/2}$ together with Eq.~(\ref{BA}) leads to a consistency condition on $s_0$,
\begin{equation}
\fracmm{28s_0^2 - 13}{3(1 + 2s_0^2)} = -1 \quad {\rm or} \quad s_0 = \sqrt{\fracmm{5}{17}} \approx 0.542 \quad {\rm and} \quad \gamma = \fracmm{8}{\sqrt{85}} \approx 0.868~.
\end{equation}
Having fixed the VEV $s_0 = \sqrt{5/17}$, we are able to calculate the mass scale $m$ and, consequently, the gravitino mass $m_{3/2}$. The numerical analysis of the inflationary dynamics in the preceding Section related the mass scale $m$ to the known amplitude $A_s$ of CMB scalar perturbations. For example, when assuming the central value,  $N=55$ with the pivot scale $k=0.05~\text{Mpc}^{-1}$ crossing the horizon, we find $m \approx 1.78 \times 10^{13}$~GeV. After substituting this value of $m$ into the vacuum relation (\ref{gravitinom}) defining the gravitino mass,  we obtain the middle value
\begin{equation} \lb{gravitinom1}
\quad m_{3/2} \approx 8.6 \times 10^{12}~\text{GeV}~.
\end{equation}

With the bare-$\mu$ solution, $s_0=\sqrt{5/17}\simeq0.542<0.668$, the decay of the inflaton into a gravitino pair is kinematically allowed. Subsequent gravitino decays can therefore contribute to the non-thermal wino abundance, although a quantitative calculation of this contribution requires information about reheating and gravitino-decay history.

Summarising the above, this Section leads to the following high-energy boundary conditions for  RG running:
\begin{equation} \label{boundary_cond}
\tan\beta=1, \quad m_0 =  m_{3/2}, \quad A = \mu_{\text{eff}} = -m_{3/2}, \quad B = -2m_{3/2}.
\end{equation}
The cosmological constraints on the soft mass parameter $m_{1/2}$ are considered in the next Section.

\begin{figure}[htbp]
    \centering
        \begin{subfigure}[b]{0.45\textwidth}
        \centering
        \includegraphics[width=\textwidth]{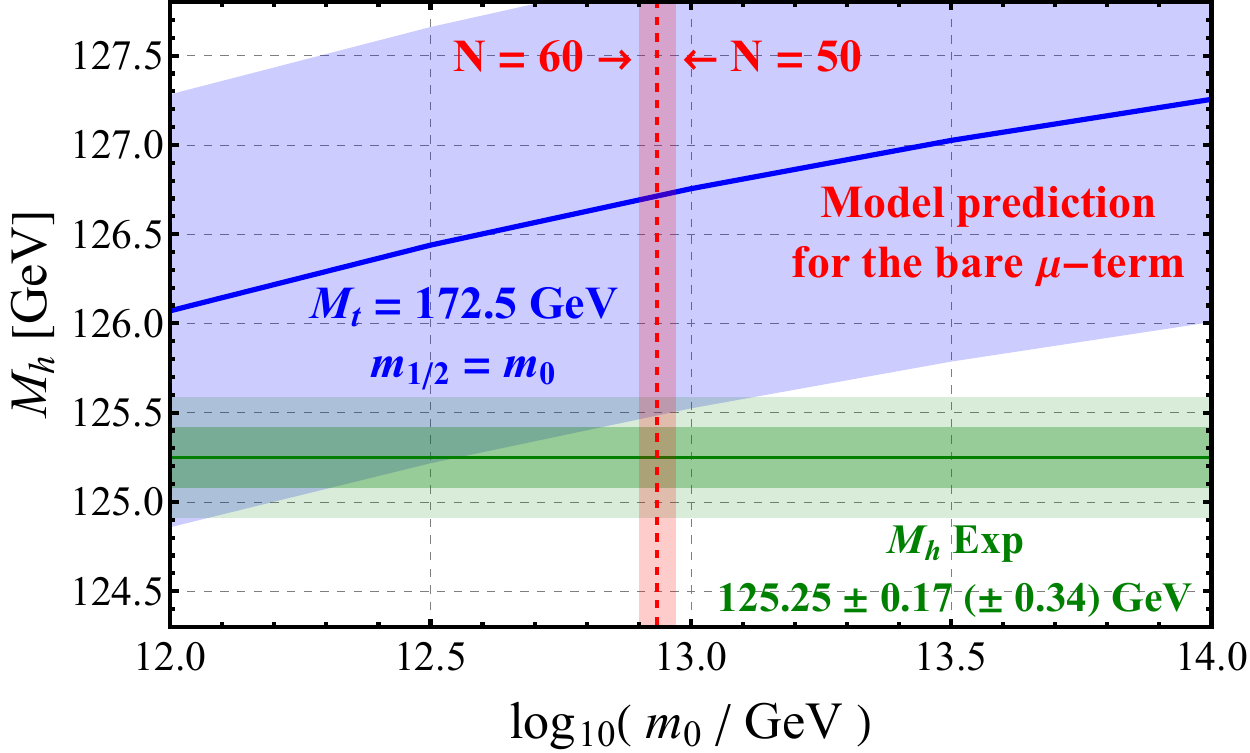}
        \caption{}
        \label{fig:R1_m0}
    \end{subfigure}
    \hfill
    \begin{subfigure}[b]{0.45\textwidth}
        \centering
        \includegraphics[width=\textwidth]{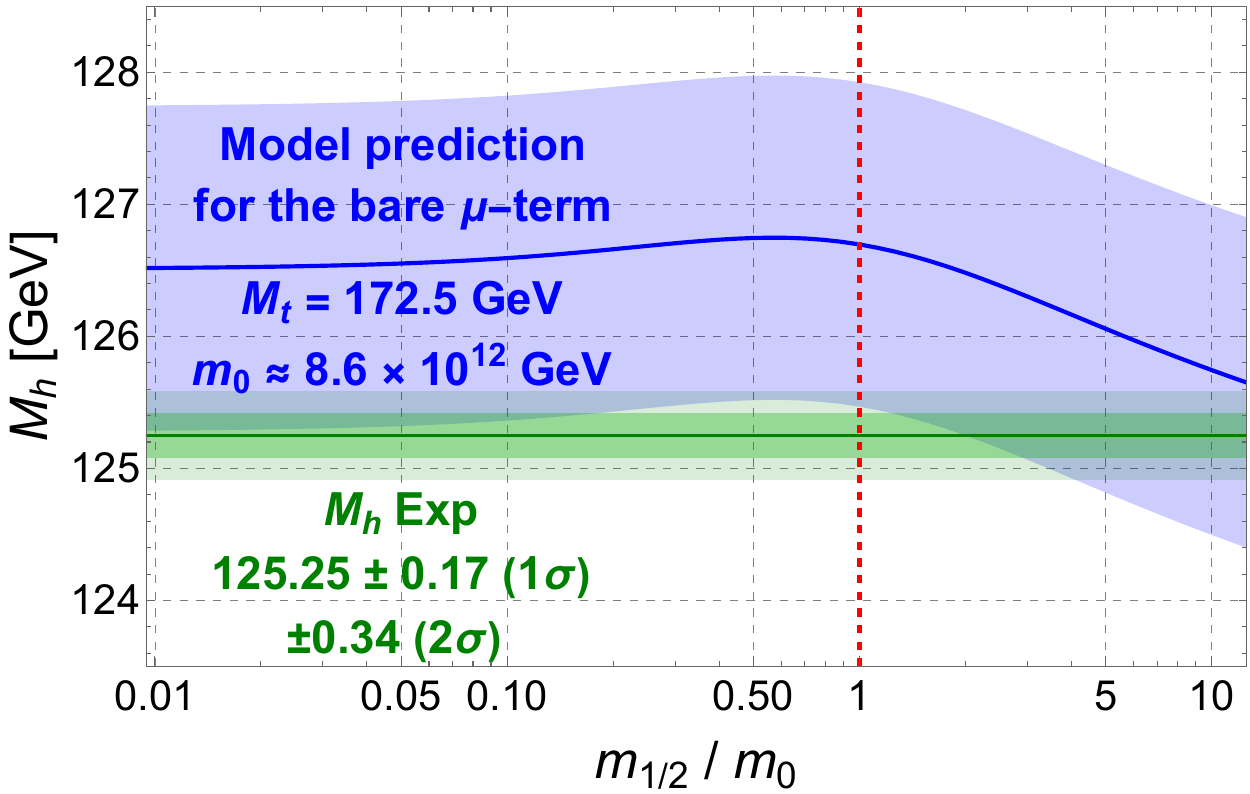} 
        \caption{}
        \label{fig:R3_m12}
    \end{subfigure}
    \caption{
    \textbf{(a)} The Higgs boson mass $M_h$ as a function of the mass scale $m_0$ with the ATLAS+CMS combined top-quark mass $M_t = 172.5$ GeV \cite{ATLAS:2023lhc} after assuming $m_{1/2} = m_0$. The red dashed line marks the model prediction for $m_0$ at $\gamma = 8/\sqrt{85}$ and $N=55$, and the solid red line shows the stability bound for $\gamma = 0.17$. The shaded red vertical band shows the variation of the $m_0$ prediction for $N \in [50, 60]$. 
    \textbf{(b)} $M_h$ as a function of the mass parameter $m_{1/2}$ for $M_t = 172.5$ GeV and the scale $m_0$ fixed by $\gamma = 8/\sqrt{85}$ and $N=55$.
    Both panels were obtained with \texttt{SusyHD}; the blue bands represent the intrinsic theory uncertainty, while the parametric uncertainty induced by $\alpha_s(M_Z)$ is not included.}
    \label{fig:higgs_mass_analysis}
\end{figure}

Our results for the Higgs boson mass $M_h$, obtained by RG evolution from the high scale ${\cal O}(10^{13})$~GeV down to the top-quark mass scale, are shown as a function of $m_0$ and $m_{1/2}$ in Fig.~\ref{fig:higgs_mass_analysis}.

As is clear from Fig.~\ref{fig:higgs_mass_analysis}, the Higgs boson mass values derived from the running are compatible with the observed Higgs mass, while these values are robust against changes in the number of e-folds $N$ and the gaugino mass scale $m_{1/2}$ for $m_{1/2} \leq m_0$.  The prediction is controlled mainly by the top-quark pole mass $M_t$ and the universal scalar mass $m_0$ that is fixed by the inflationary scale and equals the gravitino mass $m_{3/2}$. The explicit $M_t$ dependence and the associated parametric uncertainties are presented in the next Section.

We find that the Higgs boson mass rapidly approaches a plateau as $m_{1/2}$ decreases below $m_0$ and is effectively insensitive to $m_{1/2}$ throughout this region, as shown in Fig.~\ref{fig:higgs_mass_analysis}(b). The leading logarithmic effect that could significantly alter the RG running of the quartic Higgs coupling $\lambda$ is present only when both higgsinos and gauginos are light. In our framework, the higgsinos acquire masses of order the large scalar mass, and therefore decouple below the matching scale. Without light higgsinos in the EFT, the remaining light gauginos have no tree-level coupling to the Higgs boson, thus making their impact on the Higgs mass negligible. Consequently, as was also shown in Ref.~\cite{PardoVega:2015eno}, the SM alone is a reliable EFT well below the high scale $m_0$. The entire information from the SUSY spectrum becomes encoded in the threshold corrections.

When the $\mu$-term is dynamically generated by the superpotential interaction $\fracmm{\tilde{\lambda}}{M_{\text{P}}^{n-1}} \mathcal{S}^n H_u H_d$, the VEV $\VEV{\mathcal{S}} = s_0 M_{\text{P}}$ at the end of inflation yields $\mu = \tilde{\lambda} s_0^n M_{\text{P}}$. In this case, the (B)-parameter differs from that in Eq.~(\ref{BA}) as\begin{equation}\label{BA1}B = A - m_{3/2} + F^m\partial_{m}\ln \mu= A - m_{3/2} + n\fracmm{F^{\mathcal{S}}}{\mathcal{S}}= m_{3/2} \fracmm{(66 - 20n)s_0^2 + 44n - 48}{9(1 + 2s_0^2)}~.
\end{equation}
When  $n=1$, the inequality $\abs{B} \geq 2m_{3/2}$ implies  $\abs{s_0} \geq \sqrt{11/5}$ that is incompatible with 
the condition (\ref{scon}). However, when $n \geq 2$, 
Eq.~(\ref{scon}) can be satisfied. We take the minimal choice $n=2$.

Inserting Eq.~(\ref{BA1}) into the determinant condition in Eq.~(\ref{detcon1}) yields
\begin{equation} \lb{Asol_dyn}
	m_{3/2}^2 + A^2 = A \left( A - m_{3/2} + 2\fracmm{F^{\mathcal{S}}}{\mathcal{S}} \right)~.
\end{equation}
This can be rewritten as
\begin{equation}
	866 s_0^4 - 1429 s_0^2 + 527 = 0~
\end{equation}
that has two roots,
\begin{equation}
	s_0^2 = \fracmm{1429 \pm \sqrt{216513}}{1732}~.
\end{equation}
Both roots satisfy the condition (\ref{scon}) and are larger
than $0.668$.  The larger root, $s_0 \approx 1.046$, leads to higher gravitino and Higgs boson masses. Since it does not qualitatively change our conclusions, we restrict ourselves to the smaller root, $s_0 \approx 0.746$ in what follows. Then the vacuum state and the R-symmetry breaking parameter are fixed 
as

\begin{equation}
	s_0 = \sqrt{\fracmm{1429 - \sqrt{216513}}{1732}} \approx 0.746\,, \quad \gamma = \fracmm{27 \sqrt{33}+101}{\sqrt{433 \left(1429-81 \sqrt{33}\right)}} \approx 0.397~.
\end{equation}
Having fixed these parameters, the related gravitino mass for the central value, $N=55$ is given by 
\begin{equation} \lb{gravitinom2}	
m_{3/2} \approx 2.0 \times 10^{13}~\text{GeV}
\end{equation}
and differs from that in Eq.~(\ref{gravitinom1}).

Accordingly, the boundary conditions for RG running are also different, being given by
\begin{equation} \label{boundary_cond2}
	\tan\beta=1, \quad m_0 = m_{3/2}, \quad A = \mu_{\text{eff}} = \fracmm{m_{3/2}}{8} \left(9-\sqrt{33}\right), \quad B = \fracmm{m_{3/2}}{24} (63+\sqrt{33})~.
\end{equation}

The expression for $\mu_{\text{eff}}$ fixes the coupling constant $\tilde{\lambda}$ as 
\begin{equation}
 	\tilde{\lambda} = \fracmm{9-\sqrt{33}}{3} (1+2s_0^2) \fracmm{m}{M_{\text{P}}} = 36 \sqrt{\fracmm{6 \left(3431-549 \sqrt{33}\right)}{228191}} \fracmm{m}{s_0M_{\text{P}}} \approx 2.26\times 10^{-5}~.
\end{equation}

\begin{figure}[H]
    \centering
        \begin{subfigure}[b]{0.45\textwidth}
        \centering
        \includegraphics[width=\textwidth]{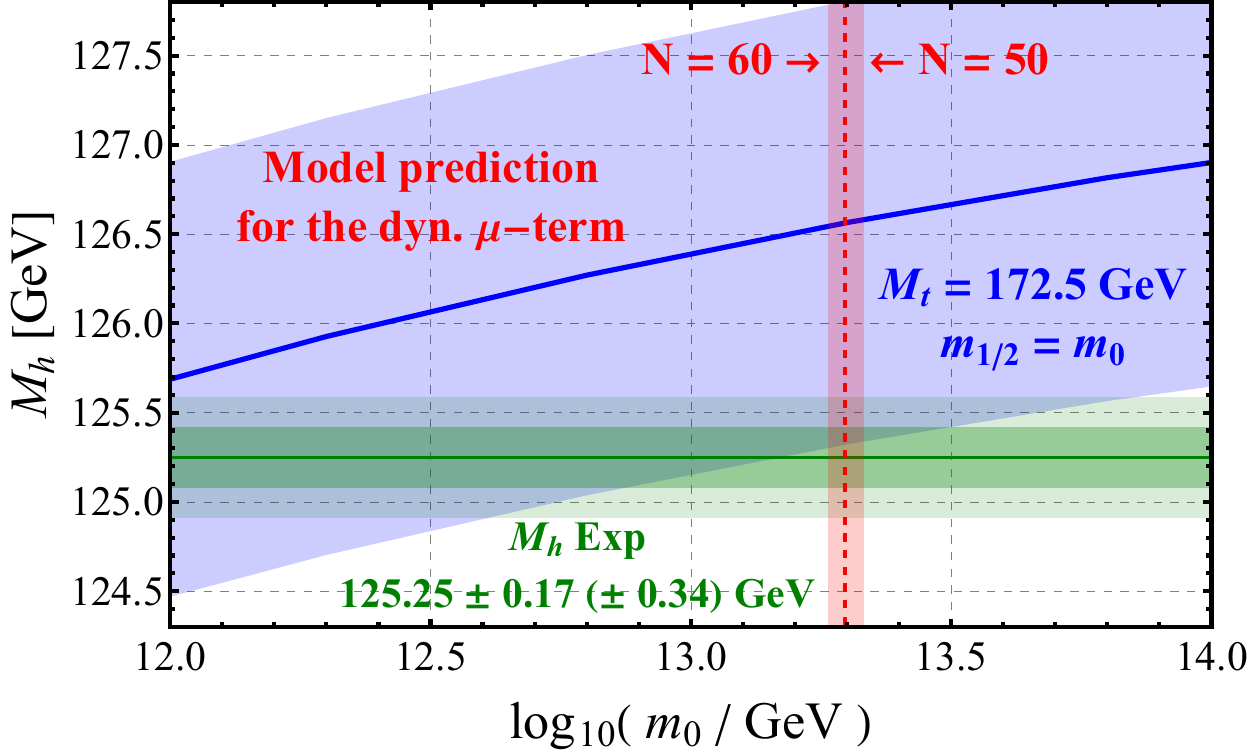}
        \caption{}
        \label{fig:R11_m0_dyn}
    \end{subfigure}
    \hfill
    \begin{subfigure}[b]{0.45\textwidth}
        \centering
        \includegraphics[width=\textwidth]{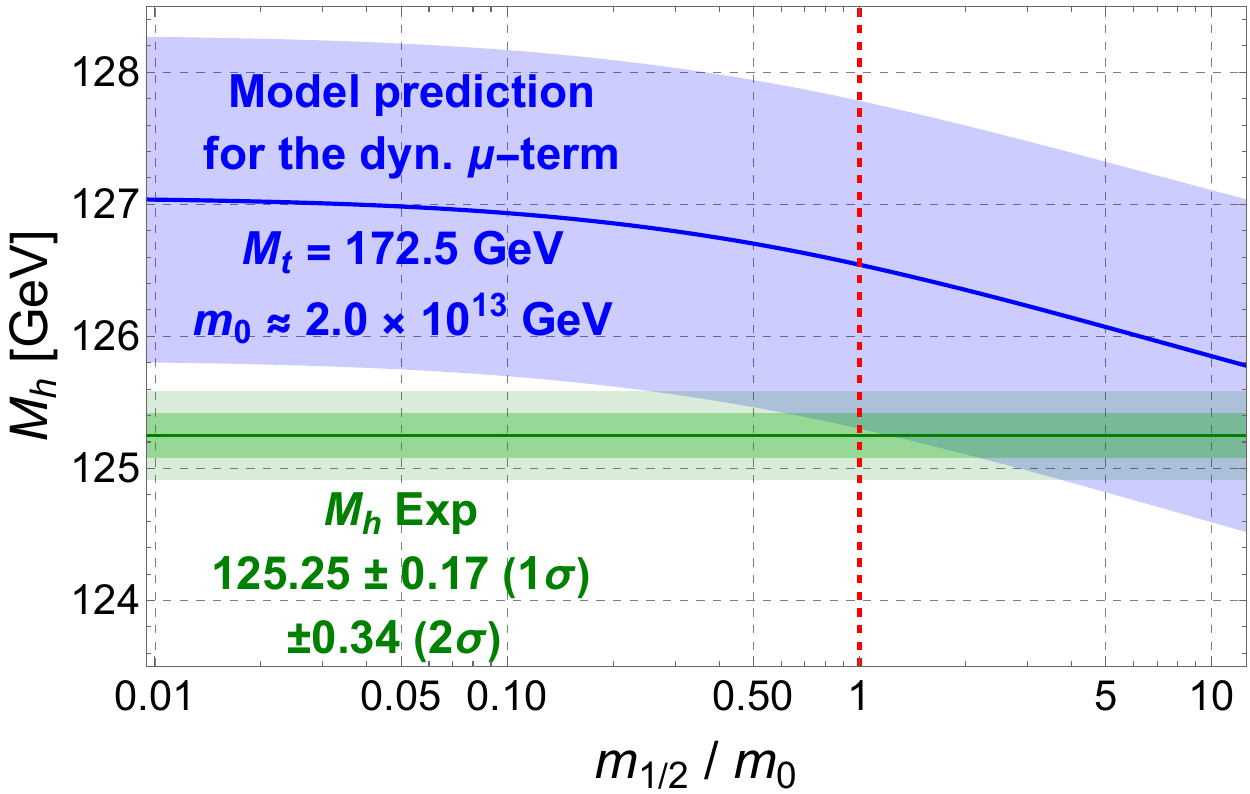} 
        \caption{}
        \label{fig:R22_m12_dyn}
    \end{subfigure}
    \caption{
    \textbf{(a)} The Higgs boson mass $M_h$ as a function of the mass scale $m_0$ with the ATLAS+CMS combined top-quark mass $M_t = 172.5$ GeV 
  \cite{ATLAS:2023lhc}
    after assuming $m_{1/2} = m_0$. The red dashed line marks the model prediction for $m_0$ at $\gamma \approx 0.397$ and $N=55$. The shaded red vertical band shows the variation of the $m_0$ prediction for $N \in [50, 60]$. 
    \textbf{(b)} $M_h$ as a function of the mass parameter $m_{1/2}$ for $M_t = 172.5$ GeV and the scale $m_0$ fixed by $\gamma \approx 0.397$ and $N=55$.
    Both panels were obtained with \texttt{SusyHD}; the blue bands represent the intrinsic theory uncertainty, while the parametric uncertainty induced by $\alpha_s(M_Z)$ is not included.}

    \label{fig:higgs_mass_analysis_dyn}
\end{figure}

Figure~\ref{fig:higgs_mass_analysis_dyn} illustrates the RG running obtained with the dynamically generated $\mu$-term. The higher scale of the gravitino mass lifts the predicted Higgs mass in the plateau region where $m_{1/2}<m_0$ relative to Fig.~\ref{fig:higgs_mass_analysis}. Nevertheless, as $m_{1/2}$ approaches $m_0$, the dynamical scenario has better agreement with the measured Higgs boson mass.

\section{Anomaly mediation of SUSY breaking and wino dark matter}
The gaugino masses $M_a$ are generated by the $F$-terms in the hidden sector and require a specification of the gauge kinetic function $f_{a}$ that is not fixed by the Starobinsky supergravity alone. In addition to the universal tree-level contributions $M_a^{\text{tree}}$, the gaugino masses receive quantum corrections via the super-Weyl anomaly. This mechanism is known as anomaly-mediated supersymmetry breaking (AMSB)~\cite{Giudice:1998xp,Moroi:1999zb,Ibe:2004tg}. Using the one-loop anomaly-induced gaugino mass~\cite{Bagger:1999rd} and taking into account the canonical K\"ahler metric in the MSSM sector, the anomalous contributions for the $U(1)_Y, SU(2)_L$, and $SU(3)_C$ gauge groups of the SM read
\begin{equation} \label{AMSB_masses}
 M_1^{\text{anom}} = \fracmm{33}{5} \fracmm{\alpha_1}{4\pi} (m_{3/2} + A)~, \quad
M_2^{\text{anom}} = \fracmm{\alpha_2}{4\pi} (m_{3/2} + 5A)~, \quad
M_3^{\text{anom}} = \fracmm{3\alpha_3}{4\pi} (A - m_{3/2})~,
\end{equation}
where $A$ is the universal trilinear soft parameter.
For a highly hierarchical spectrum, we should also include finite threshold corrections arising from integrating out the heavy Higgs and higgsino states at the matching scale $m_0$~\cite{Bhattacherjee:2012ed},
\begin{equation} \label{Higgs_thresholds}
 \Delta M_1^H = \fracmm{3}{5} \fracmm{\alpha_1}{4\pi} L~, \quad
 \Delta M_2^H = \fracmm{\alpha_2}{4\pi} L~, \quad
 \Delta M_3^H = 0~,
\end{equation}
where the loop function $L$ of the $\mu_{\text{eff}}$ parameter is \cite{Bhattacherjee:2012ed}
\begin{equation}
    L = \mu_{\text{eff}} \sin 2\beta \fracmm{m_A^2}{\abs{\mu_{\text{eff}}}^2 - m_A^2} \ln \fracmm{\abs{\mu_{\text{eff}}}^2}{m_A^2}~~.
\end{equation}
 With the high-scale boundary conditions $\sin 2\beta = 1$ and $m_A^2 = 2m_{3/2}^2 + 2\abs{\mu_{\text{eff}}}^2$, the sign of $L$ is determined by the sign of $\mu_{\text{eff}}$. With our parameters, the threshold corrections are comparable to the AMSB contributions.

Hence, the total gaugino mass at the matching scale is a sum of the tree-level contribution, the quantum AMSB contribution, and the heavy-Higgs threshold correction,
\begin{equation} \label{M_total}
 M_a^{\text{total}} = M_a^{\text{tree}} + M_a^{\text{anom}} + \Delta M_a^H~.
\end{equation}
 The high-scale spectrum strongly restricts the neutralino LSP. Higgsinos are heavy because $\abs{\mu_{\text{eff}}}\sim m_{3/2}$. In the minimal thermal history, a nearly pure bino is not viable because, once sfermions and higgsinos are decoupled, its annihilation rate is too small to reproduce the observed relic abundance. Non-minimal bino scenarios involving co-annihilation, resonant annihilation, late entropy dilution or R-parity violation are outside the scope of this investigation. A nearly pure wino, by contrast, retains unsuppressed $SU(2)_L$ annihilation and chargino--neutralino co-annihilation, so we focus on the wino LSP. Since the tree-level contribution in Eq.~(\ref{gkf}) is universal, the relative splittings amongst the total gaugino masses are controlled by the quantum contribution. The thermal wino is realised by choosing the universal tree-level contribution so that it cancels most of the quantum contribution to $M_2$, leaving TeV-scale winos while binos and gluinos remain much heavier. Constraints on $M_a^{\text{total}}$ can be inferred from the Higgs boson mass running in Figs.~\ref{fig:higgs_mass_analysis}(b) and \ref{fig:higgs_mass_analysis_dyn}(b). In the standard thermal scenario, the relic-density condition fixes the canonical physical wino target mass, as discussed below.

 In our setup, the MSSM scalars are lighter than the inflaton.
Consequently, these scalars can be efficiently produced by gravitational particle
production at the end of inflation~\cite{Kolb:2023ydq}. Following
Ref.~\cite{Ema:2018ucl}, the present-day energy density of particular  gravitationally
produced scalar species, normalised by the entropy density $s$, scales as
\begin{equation}
    \fracmm{\rho_0^{(\mathrm{GP})}}{s}
    \sim
    \fracmm{\mathcal{C}}{4}
    \fracmm{m_{0} H_{\mathrm{inf}} T_{\mathrm{R}}}{M_P^2}
    \simeq
    4 \times 10^{-10}~\mathrm{GeV}\cdot \mathcal{C}
    \left(\fracmm{m_{0}}{10^9~\mathrm{GeV}}\right)
    \left(\fracmm{H_{\mathrm{inf}}}{10^9~\mathrm{GeV}}\right)
    \left(\fracmm{T_{\mathrm{R}}}{10^{10}~\mathrm{GeV}}\right).
\end{equation}
Here $\mathcal{C}\sim 10^{-2}$ and $T_{\mathrm{R}}$ is the reheating temperature,
typically of order $10^8$--$ 10^9~\mathrm{GeV}$~\cite{Ema:2024sit}. With
$H_{\mathrm{inf}}\sim \mathcal{O}(10^{13})~\mathrm{GeV}$ and
$m_0\sim \mathcal{O}(10^{13})~\mathrm{GeV}$, this estimate implies that a stable scalar
population can exceed the observed DM abundance by many orders of magnitude.
The precise abundance depends on a production coefficient, a reheating history and a
number of gravitationally produced species. Avoiding the scalar overproduction requires the gravitationally produced scalars to be
unstable. A possible way to realise that is to make wino the LSP, so that the heavy
scalars decay into final states containing winos. In that case the scalar overabundance is
not automatically inherited by the wino LSP. Neglecting subsequent wino annihilation and
entropy dilution, the injected wino yield is schematically given by
\begin{equation}
    Y_{\tilde W}^{\mathrm{inj}}
    \sim
    N_{\tilde W}\,
    \mathrm{Br}(\phi\to \tilde W+X)\,
    \fracmm{\rho_\phi^{(\mathrm{GP})}}{m_0 s},
\end{equation}
where $N_{\tilde W}$ is the average number of winos produced per scalar decay and
$\mathrm{Br}(\phi\to \tilde W+X)$ denotes the relevant branching fraction. The
corresponding relic abundance is then estimated as
\begin{equation}
    \Omega_{\tilde W}h^2
    \sim
    \fracmm{M_2}{m_0}\,
    N_{\tilde W}\,
    \mathrm{Br}(\phi\to \tilde W+X)\,
    \fracmm{\rho_\phi^{(\mathrm{GP})}}{s}
    \fracmm{s_{\rm today}}{\rho_c/h^2}.
\end{equation}
Hence, lowering the ratio $M_2/m_0$ can reduce the final energy density transferred to
the wino LSP. However, this estimate is not sufficient to determine the final relic
abundance. In the non-thermal regime
\begin{equation}
    m_0 > M_2 > T_{\mathrm R},
\end{equation}
thermal wino production is Boltzmann suppressed, and the final abundance depends on the
post-inflationary history, including the scalar yield, decay temperature, branching ratios
and multiplicities, possible wino annihilation after injection, and any late-time entropy
dilution. A reliable assessment of this branch therefore requires a dedicated calculation,
which we leave for future work.

A qualitatively different possibility is to lower the gaugino masses so that the wino mass
lies well below the reheating temperature, $M_2 \ll T_{\mathrm{R}}$. In this thermal regime,
the unsuppressed $SU(2)_L$ gauge interactions efficiently bring winos into chemical and
kinetic equilibrium with the SM plasma. The final relic abundance is then determined by
standard thermal freeze-out and is insensitive to the initial gravitational-production
yield~\cite{Moroi:1999zb}. This should be distinguished from the non-thermal regime
$M_2 > T_{\mathrm{R}}$, where the final abundance depends on the post-inflationary history,
including gravitational production, decay chains, possible annihilation after injection,
and entropy dilution.

This corresponds to the thermal ``wino miracle'' scenario. For a pure wino, the neutral and
charged components form an almost degenerate $SU(2)_L$ multiplet, so co-annihilation with
the charged wino states is essential in the relic density calculation. In addition, for
TeV-scale electroweak multiplets the exchange of electroweak gauge bosons generates a
long-range potential between the slowly moving annihilating particles. The resulting
Sommerfeld enhancement substantially increases the effective annihilation cross section at
freeze-out~\cite{Hisano:2006nn,Hryczuk:2010zi,Beneke:2014gja,Beneke:2016ync}. Established
calculations including these effects place the thermal  wino mass at about
$2.7$--$3.0$ TeV, with a mild dependence on higher-order electroweak corrections and possible
departures from the pure-wino limit~\cite{Hisano:2006nn,Beneke:2016ync,Beneke:2020vff}.
We therefore adopt $M_2\simeq3$ TeV as the thermal target and do not recalculate the
freeze-out abundance in this work.

Direct detection constraints are also mild for the pure wino limit realised here. Since the
Higgsino mass parameter is of  the order of the high SUSY-breaking scale,
$\mu_{\rm eff} \sim 10^{13}$ GeV, the Higgsino admixture of the neutral wino is suppressed by
$m_W/\mu_{\rm eff}$ and the tree-level Higgs-mediated spin-independent scattering amplitude is
negligible. The tree-level $Z$-exchange contribution is absent for a Majorana neutral
wino. The leading spin-independent scattering therefore arises from EW loop effects.
One-loop diagrams generate effective couplings of the wino to quarks and to the Higgs
sector, while the effective coupling to gluons first appears at two loops and gives an
important contribution to the nucleon matrix element~\cite{Hisano:2004pv,Hisano:2010ct,Hisano:2015rsa}.
For a nearly pure TeV-scale wino, the spin-independent wino-proton cross section is only
weakly dependent on the wino mass and is predicted to be
\begin{equation}
\sigma_{\rm SI}^{p} \simeq 2.3\times 10^{-47}\ {\rm cm}^2
\end{equation}
with an uncertainty of order $(30$--$40)\%$ from perturbative and hadronic inputs~\cite{Hisano:2015rsa}.
This value is below current LZ bounds for multi-TeV dark matter, but it lies above the
irreducible neutrino background and is within the target range of next-generation multi-ton
liquid-xenon experiments such as DARWIN/XLZD~\cite{LZ:2024lux,XLZD:2024,DARWIN:2016hyl}.
 Thus, direct detection alone does not exclude the $2.7$--$3.0~{\rm TeV}$ pure-wino state, and next-generation multi-ton liquid-xenon experiments can directly probe this scenario.

Indirect detection gives a qualitatively stronger constraint on its interpretation as all of the DM. A recent dedicated analysis of 14 years of Fermi-LAT data excludes a pure thermal wino constituting $100\%$ of the DM under the standard cosmological history~\cite{Safdi:2025sfs}. This conclusion was found to persist for a range of Galactic DM density profiles, including conservative cored profiles. The statement is conditional on the astrophysical assumptions and on the standard cosmological history used in that analysis, but within those assumptions the canonical $2.7$--$3.0~{\rm TeV}$ thermal wino cannot constitute all of the observed DM.

The wino state itself is not excluded as a particle. Since the annihilation signal scales approximately with the square of the wino DM fraction, a subdominant wino component can evade these bounds. A non-standard cosmological history can also alter the relation between the wino mass and its present abundance. A quantitative treatment of these possibilities is beyond the scope of this work. We therefore use the $2.7$--$3.0~{\rm TeV}$ mass as the canonical thermal-wino benchmark for the particle-physics analysis, without assuming that this state constitutes all of the present DM in the standard cosmological history.


 To realise the canonical $M_2\simeq3~{\rm TeV}$ wino benchmark,
we tune the universal tree-level contribution  
$M_a^{\text{tree}}$ in order to cancel the large quantum 
corrections in the $SU(2)_L$ sector. 

In the bare $\mu$-term case, the loop function is negative, 
$L < 0$. We fix the universal tree-level gaugino mass as 
$M_a^{\text{tree}} \approx 1.06 \times 10^{11}$~GeV. 
This yields the following total gaugino mass spectrum 
at the high scale:
\begin{equation}
 M_1^{\text{total}} \approx 8.83 \times 10^{10}\text{ GeV}, 
\quad
 M_2^{\text{total}} \approx 1.63 \times 10^{3}\text{ GeV}, 
\quad
 M_3^{\text{total}} \approx -1.09 \times 10^{10}\text{ GeV}.
\end{equation}

In the dynamically generated $\mu$-term case, 
the loop function is positive, $L > 0$. 
We fix the tree-level mass as
 $M_a^{\text{tree}} \approx -1.73 \times 10^{11}$~GeV, 
which leads to
\begin{equation}
M_1^{\text{total}} \approx 2.02 \times 10^{11}\text{ GeV}, 
\quad
M_2^{\text{total}} \approx 1.61 \times 10^{3}\text{ GeV}, 
\quad
M_3^{\text{total}} \approx -2.50 \times 10^{11}\text{ GeV}.
\end{equation} 
In both cases, the wino mass $M_2^{\text{total}}$ is fixed at the matching scale $m_0$ and runs downwards to yield $M_2 \approx 3$~TeV at the low-energy scale. We calculate the Higgs mass with two independent EFT implementations: \texttt{SusyHD}~\cite{PardoVega:2015eno}~\footnote{
While comparing \texttt{SusyHD} and \texttt{HSSUSY} predictions, we found a bug in \texttt{SusyHD}: the fast-interpolation table distributed with \texttt{SusyHD} v1.0.2 was independent of its $g_3$ coordinate and therefore missed part of the $\alpha_s$ dependence. We regenerated the table with the correct $g_3$ boundary condition and used this local corrected version throughout the analysis. The correction leaves the central Higgs masses and intrinsic \texttt{SusyHD} theory uncertainties unchanged at the quoted precision, but restores the correct $\alpha_s$ uncertainty. Appendix~\ref{app:higgs_uncertainties} gives the complete comparison, explains the uncertainty prescriptions and documents the correction. The corrected \texttt{SusyHD} source, regenerated table, regeneration script and README file are provided as ancillary files with the arXiv submission.}  and \texttt{HSSUSY} within \texttt{FlexibleSUSY}~\cite{Athron:2016fuq,Athron:2017fvs}.
Both packages match the MSSM directly onto the SM at the high SUSY scale and
re-sum the large logarithms generated by the separation between this scale and
the electroweak scale. We use \texttt{SusyHD} for the main numerical analysis and
\texttt{HSSUSY} as an independent cross-check of the Higgs mass predictions and their
uncertainty estimates.

 We find that both EFT implementations give a Higgs boson mass compatible with the measured value after imposing the thermal wino condition. With the central input values $M_t=172.52~{\rm GeV}$ and
$\alpha_s(M_Z)=0.1180$, we obtain

\begin{equation}
\begin{aligned}M_h^{\rm \texttt{SusyHD}}({\rm bare}\ \mu)&=126.85\pm1.23_{\rm th}\pm0.64_{\alpha_s}~{\rm GeV},
\\
M_h^{\rm \texttt{HSSUSY}}({\rm bare}\ \mu)&=126.31\pm0.53_{\rm th}\pm0.64_{\alpha_s}~{\rm GeV},
\\
M_h^{\rm \texttt{SusyHD}}({\rm dynamical}\ \mu)&=127.36\pm1.23_{\rm th}\pm0.64_{\alpha_s}~{\rm GeV},
\\
M_h^{\rm \texttt{HSSUSY}}({\rm dynamical}\ \mu)&=126.82\pm0.42_{\rm th}\pm0.65_{\alpha_s}~{\rm GeV}.\end{aligned}
\end{equation}

The \texttt{SusyHD} and \texttt{HSSUSY} central predictions differ by about $0.5~{\rm GeV}$ for both the bare and dynamical $\mu$ scenarios, and are consistent within the quoted theory and parametric uncertainties.

The two quoted errors have different origins. The theory error estimates the
missing higher-order contributions. \texttt{SusyHD} provides separate low-energy SM,
high-scale SUSY-matching and power-suppressed EFT components, which we combine
linearly.

For the bare $\mu$ case, this gives

\begin{equation}
\Delta_{\rm th}M_h^{\rm \texttt{SusyHD}}
=
1.22813+0.00138+0.00057
=
1.23008~{\rm GeV},
\end{equation}

while for the dynamically generated $\mu$ case, we get

\begin{equation}
\Delta_{\rm th}M_h^{\rm \texttt{SusyHD}}
=
1.22759+0.00219+0.00046
=
1.23024~{\rm GeV}.
\end{equation}

\begin{figure}[H]
    \centering
    \begin{subfigure}[b]{0.47\textwidth}
        \centering
        \includegraphics[width=\textwidth]{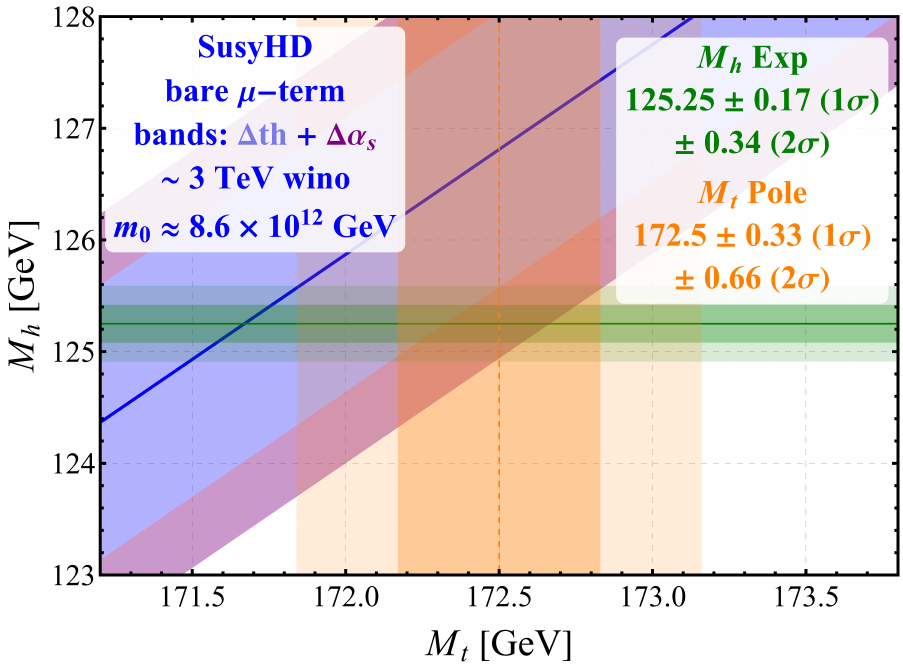}
        \caption{\texttt{SusyHD}, bare $\mu$ term}
        \label{fig:hmass_SusyHD_baremu}
    \end{subfigure}
    \hfill
    \begin{subfigure}[b]{0.47\textwidth}
        \centering
        \includegraphics[width=\textwidth]{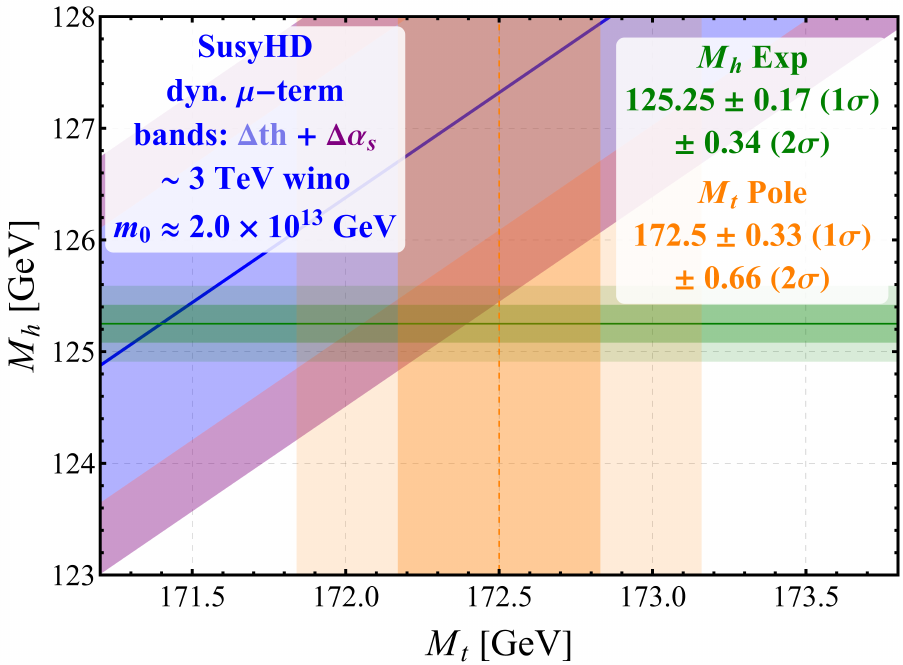}
        \caption{\texttt{SusyHD}, dynamically generated $\mu$ term}
        \label{fig:hmass_SusyHD_dmu}
    \end{subfigure}
    
    \vspace{0.3cm}
    
    \begin{subfigure}[b]{0.47\textwidth}
        \centering
        \includegraphics[width=\textwidth]{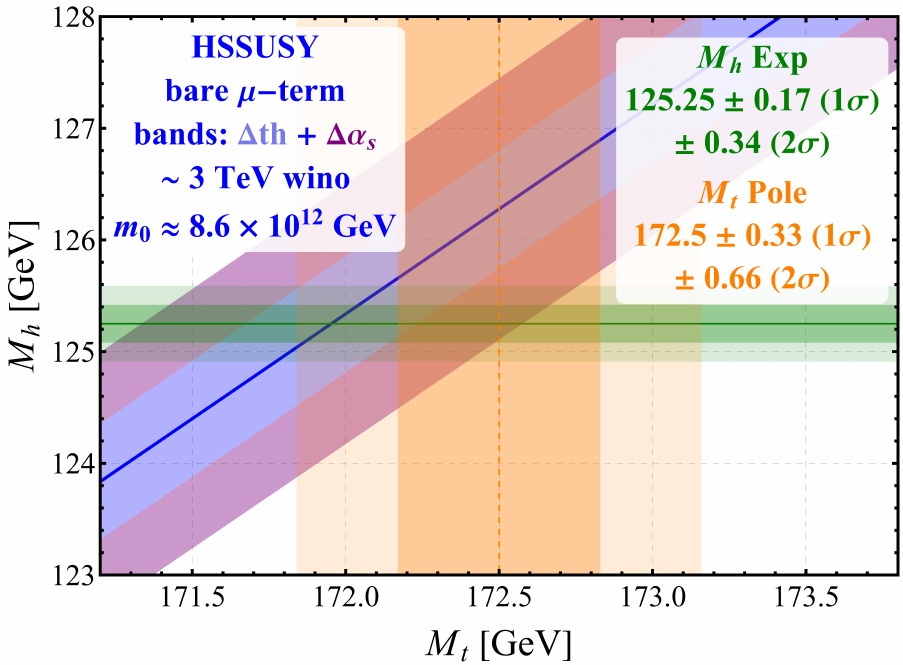}
        \caption{\texttt{HSSUSY}, bare $\mu$ term}
        \label{fig:hmass_HSSUSY_baremu}
    \end{subfigure}
    \hfill
    \begin{subfigure}[b]{0.47\textwidth}
        \centering
        \includegraphics[width=\textwidth]{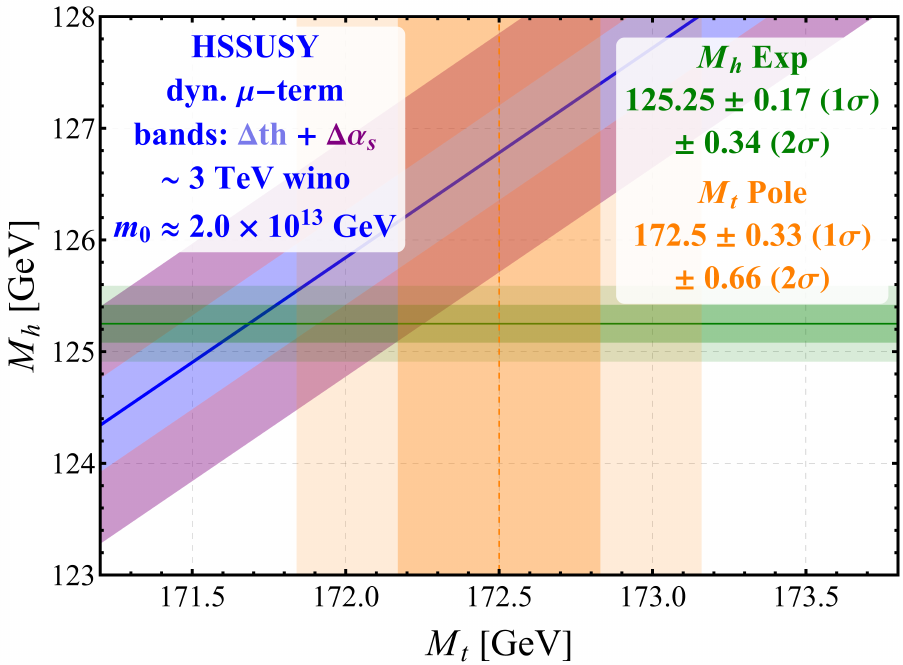}
        \caption{\texttt{HSSUSY}, dynamically generated $\mu$ term}
        \label{fig:hmass_HSSUSY_dmu}
    \end{subfigure}
    \caption{
    Higgs boson mass $M_h$ as a function of the ATLAS+CMS combined top-quark pole mass $M_t$~\cite{ATLAS:2023lhc}. 
    Panels (a) and (c) correspond to the bare-$\mu$ scenario with $\gamma=8/\sqrt{85}$, while panels (b) and (d) correspond to the dynamically generated $\mu$ scenario with $\gamma\simeq 0.397$. 
    In all panels, the high-scale total gaugino masses are chosen so that RG evolution gives a physical low-energy wino mass $M_2\simeq3~{\rm TeV}$. 
    The green shaded regions show the $1\sigma$ and $2\sigma$ experimental constraints on $M_h$, while the orange shaded regions show the corresponding constraints on $M_t$ from the Particle Data Group (PDG) 2024~\cite{ParticleDataGroup:2024cfk}. 
    The blue bands show the intrinsic theory uncertainty $\Delta_{\rm th}M_h$, while the purple extensions show the additional parametric uncertainty induced by $\alpha_s(M_Z)$. The two uncertainties are displayed additively for visual comparison but are quoted separately in the text.
    }
    \label{fig:higgs_mass_SusyHD_HSSUSY_comparison}
\end{figure}

Therefore, the \texttt{SusyHD} theory error is almost entirely the low-energy SM component  dominated by missing higher-order terms in the extraction of the running top Yukawa coupling and in the conversion of the running Higgs quartic coupling into the Higgs pole mass. \texttt{HSSUSY} uses a more recent prescription based on variations of the higher-order terms entering the extraction of the running top Yukawa coupling, the renormalisation scale in the Higgs pole mass calculation and the high-scale matching. It gives theory uncertainties of $ 0.5264~{\rm GeV}$ and $ 0.4235~{\rm GeV}$ for the bare and dynamically generated $\mu$ cases, respectively. The different theory errors therefore reflect different uncertainty prescriptions rather than a disagreement between the central predictions~\cite{Slavich:2020zjv}.

The $\alpha_s$ error is parametric. Varying $\alpha_s(M_Z)=0.1180\pm0.0009$ changes the QCD conversion of the top pole mass into the running top Yukawa coupling and, hence, the RG evolution of $y_t$ and $\lambda$ over the large interval between the electroweak and SUSY scales. The resulting uncertainties are $ 0.6406~{\rm GeV}$ and $ 0.6428~{\rm GeV}$ in \texttt{SusyHD}, and $ 0.6442~{\rm GeV}$ and $0.6464~{\rm GeV}$ in \texttt{HSSUSY}, for the bare and dynamically generated $\mu$ scenarios, respectively. At fixed $M_t$, we add the theory and $\alpha_s$ uncertainties linearly, giving total vertical uncertainties of $ 1.87~{\rm GeV}$ and $ 1.17~{\rm GeV}$ for the bare-$\mu$ \texttt{SusyHD} and \texttt{HSSUSY} results, and $ 1.87~{\rm GeV}$ and $ 1.07~{\rm GeV}$ for the corresponding dynamical-$\mu$ results. This linear sum is a conservative uncertainty envelope, not a statistical confidence interval. The top-mass uncertainty, $\Delta_{M_t}M_h\simeq 0.62~{\rm GeV}$, is kept
separate because $M_t$ is varied explicitly along the horizontal axis in
Fig.~\ref{fig:higgs_mass_SusyHD_HSSUSY_comparison}.

The resulting thermal wino scenario is sharply predictive at future colliders. The
robustness of this prediction originates from the highly hierarchical SUSY spectrum
realised in our framework. The Higgsino mass parameter, together with the bino and gluino
masses, resides at the scale
$\mu,\; M_1,\; M_3 \sim 10^{10}$--$10^{13}\ {\rm GeV}$,
whereas the wino mass is of order $M_2 \simeq 3~{\rm TeV}$. The chargino mass matrix contains
off-diagonal entries of order $m_W$, while the Higgsino mass parameter is near the high
supersymmetry-breaking scale. Consequently, the Higgsino admixture in the light chargino and
neutralino eigenstates is suppressed by $m_W/\mu$, and the corresponding mass corrections
are negligible. The lightest chargino and neutralino are therefore effectively pure wino
states.

Since the charged and neutral winos belong to the same $SU(2)_L$ multiplet, they are degenerate at tree level in the pure-wino limit. Two-loop electroweak corrections generate a mass splitting of about $160$--$165~{\rm MeV}$ for a multi-TeV wino, giving a chargino lifetime of approximately $0.2~{\rm ns}$ and, hence,
$$
c\tau\simeq6~{\rm cm}.
$$
The dominant decay is $\tilde{\chi}^{\pm}_1\to\tilde{\chi}^{0}_1\pi^\pm$, and the pion is usually too soft to be reconstructed~\cite{Ibe:2012sx}. In the laboratory frame, the mean flight distance is $\beta\gamma c\tau$, so boosted charginos can cross several inner tracking layers before decaying and produce a reconstructable disappearing track. The physical lifetime of $0.2~{\rm ns}$ is somewhat shorter than the lifetime near $1~{\rm ns}$ for which the ATLAS sensitivity is maximal, because a larger fraction of the longer-lived charginos leave enough inner-layer hits. Nevertheless, it is well within the effective search region. Using tracklets reconstructed from three or four innermost-layer measurements, the latest ATLAS analysis excludes the theoretical pure-wino line up to about $670~{\rm GeV}$; allowing the lifetime to vary, the strongest observed wino limit is about $880~{\rm GeV}$ near $1~{\rm ns}$~\cite{ATLAS:2026dt}.

At the HL-LHC, with $\sqrt{s}=14~{\rm TeV}$ and $3~{\rm ab}^{-1}$, the ATLAS disappearing-track projection gives a $5\sigma$ discovery reach of about $0.8~{\rm TeV}$ and a $95\%$ confidence-level exclusion reach of about $1.1~{\rm TeV}$ for a pure wino~\cite{ATLAS:2018hllhcDT}. The much larger data set will therefore extend the present sensitivity substantially, but the limited centre-of-mass energy prevents the HL-LHC from reaching the thermal wino mass.
 The HL-LHC will provide an important indirect test of the scenario through improved precision on the SM inputs entering the Higgs mass prediction. The combined ATLAS+CMS HL-LHC projection gives an expected Higgs boson mass uncertainty of about $21~{\rm MeV}$, while the top-quark mass uncertainty could reach about $200~{\rm MeV}$ in optimistic profiled analyses, with more conservative projections at the few-hundred-MeV level~\cite{ATLASCMS:2025hllhcproj,CMS:2025topmassHL}. Since the predicted Higgs mass in our framework is sensitive to $M_t$ and the high-scale threshold corrections fixed by the SUSY-breaking sector, those measurements will significantly reduce the allowed parameter space and provide a complementary indirect probe of our scenario.

A future $100~{\rm TeV}$ proton collider changes the situation qualitatively. Dedicated disappearing-track studies show that its discovery reach extends beyond the canonical thermal-wino mass scale, $m_{\tilde{\chi}^{\pm}1}\simeq m{\tilde{\chi}^{0}_1}\simeq 3~{\rm TeV}$, for tracker configurations optimised for short charged-particle tracks~\cite{Saito:2019rtg}. This thermal wino  may still be realised as a subdominant DM component or in a non-standard cosmological history. Therefore, a $100~{\rm TeV}$ proton collider can directly test the  pure-wino state through disappearing tracks, independently of its cosmological abundance.


\section{Conclusion}

In this paper, we proposed a framework connecting cosmic inflation and CMB observables to particle physics phenomenology at colliders and direct detection experiments, with SUSY providing the link between them. In our approach, the hidden sector is not introduced {\it ad hoc} but emerges from the dual description of the Starobinsky supergravity that gravitationally couples to the MSSM in the Einstein frame. Inflation is realised in the bosonic part of the hidden sector and leads to a Minkowski vacuum with a non-vanishing VEV $s_0$ of the real scalar component of the goldstino superfield $\mathcal{S}$.

The inflation sector is effectively controlled by two parameters, the mass parameter $m$ and the vacuum expectation value $s_0$.  The parameter $m$ is fixed by the CMB amplitude of the scalar power spectrum and the number of e-folds between horizon crossing at the pivot scale and the end of inflation. The non-vanishing $s_0$ triggers spontaneous SUSY breaking, with the gravitino mass $m_{3/2}$ determined by $m$ and $s_0$.  For all values of $s_0$ considered in this paper, the inflationary dynamics is effectively of the single-field type. The predicted scalar tilt $n_s$ is slightly larger than in the original Starobinsky model, thus improving agreement with the recent ACT data.

Spontaneous SUSY breaking is gravitationally mediated to the MSSM, whose soft parameters are derived in terms of $m_{3/2}$ and $s_0$. The same hidden sector can also generate a high-scale Higgs sector $\mu$ term. In particular, the coupling $\mathcal{S}^2 H_u H_d$ can realise a singlet-generated $\mu$ mechanism, where the singlet is identified with the goldstino superfield already present in Starobinsky supergravity. Hence, no additional singlet has to be introduced ``by hand''. We also considered the case in which the MSSM contains a bare $\mu$ term independent of the hidden sector.

To connect the resulting MSSM spectrum to SM phenomenology, we demand the existence of a light SM Higgs doublet. This implies that the determinant of the Higgs-sector mass-squared matrix should vanish at the high scale, $\det(\mathcal{M}_H^2)=0$. With minimal stop mixing, this gives an equation on $s_0$. Its solution fixes the inflationary predictions, the gravitino mass and the soft parameters,  except for the gaugino masses because the gauge kinetic function is not specified by Starobinsky supergravity alone. Instead, the gaugino sector is constrained by the post-inflationary and DM phenomenology.

With conserved R-parity, the LSP is stable and can constitute DM. Since the MSSM scalars are lighter than the inflaton, they can be gravitationally produced after inflation and, depending on the reheating history, their population can exceed the observed DM abundance if they are stable. A gaugino LSP avoids a stable heavy-scalar relic. After including quantum contributions, the LSP can be a neutral wino. The wino abundance can be either non-thermal or thermal. A reliable calculation of the non-thermal abundance requires a detailed treatment of reheating, heavy-particle decays, subsequent annihilation and possible entropy production, which we leave for future work. In the standard thermal scenario, the relic-density calculation selects a nearly pure wino with a physical mass around $3~{\rm TeV}$, but realising this  requires a high-precision cancellation between the universal tree-level and quantum contributions to $M_2$. Moreover, a recent dedicated analysis of Fermi-LAT gamma-ray data excludes this pure thermal wino as the sole DM component under the standard cosmological history, including for conservative cored Galactic profiles~\cite{Safdi:2025sfs}. Viable wino cosmologies therefore require either a subdominant wino component or a non-standard cosmological history.

The resulting spectrum with scalar superpartners and higgsinos near the inflationary scale and a TeV-scale wino LSP is characteristic of split-SUSY scenarios~\cite{Arkani-Hamed:2004zhs}. The soft terms determine the high-scale threshold corrections and the boundary condition for the RG evolution of the Higgs quartic coupling, while its tree-level MSSM contribution vanishes for $\tan\beta=1$. Three-loop RG evolution down to the top-quark mass scale gives a Higgs boson mass compatible with experiment. Independent calculations with \texttt{SusyHD} and \texttt{HSSUSY} agree on the central prediction at the level of about $0.5~{\rm GeV}$. We quote the intrinsic theory uncertainty separately from the parametric uncertainty induced by $\alpha_s(M_Z)$ and adopt the more recent \texttt{HSSUSY} estimate for the residual perturbative uncertainty. This establishes the link between the CMB normalisation, the scale of SUSY breaking and the Higgs-boson mass. The predicted Higgs mass depends only weakly on the number of e-folds and becomes effectively insensitive to $m_{1/2}$ once $m_{1/2}<m_0$.

A pure wino with a mass around $3~{\rm TeV}$ remains beyond the reach of the HL-LHC. Electroweak radiative corrections split the charged and neutral wino states by about $160~{\rm MeV}$, giving a chargino lifetime of about $0.2~{\rm ns}$ and the characteristic disappearing-track signature. For $\sqrt{s}=14~{\rm TeV}$ and $3~{\rm ab}^{-1}$, HL-LHC projections give a discovery reach of about $0.8~{\rm TeV}$ and an exclusion reach of about $1.1~{\rm TeV}$ for a pure wino~\cite{ATLAS:2018hllhcDT}, while a future $100~{\rm TeV}$ proton collider is projected to reach the $3~{\rm TeV}$ mass scale. The same state is compatible with existing direct-detection bounds: tree-level spin-independent scattering is suppressed by the very heavy higgsino mass, while the loop-induced wino-nucleon cross section lies below current sensitivities but within the projected reach of next-generation multi-ton liquid-xenon detectors. Direct detection and disappearing-track searches therefore provide complementary tests of the pure-wino particle state, independently of its cosmological abundance, while its viability as the sole  DM  component is separately constrained by indirect detection.

The same strategy can be applied to other realisations of inflation and SUSY breaking in supergravity. Different K\"ahler potentials and superpotentials lead to different soft terms and, hence,  different high-scale boundary conditions for RG evolution. One may also consider coupling Starobinsky supergravity to the MSSM in the Jordan frame or generating the $\mu$ term through a combination of bare and dynamical contributions. The master $f$-function can be modified to allow primordial black hole production providing the additional non-particle DM candidate~\cite{Aldabergenov:2020bpt,Aldabergenov:2020yok,Ishikawa:2021xya}, which makes pure wino LSP the subdominant DM component.

 Our framework can naturally accommodate neutrino masses through an embedding of the visible sector into a supersymmetric $SO(10)$ grand unified theory. In this case, each matter generation, including a right-handed neutrino, is contained in a single $16$-plet  representation, while $B-L$ breaking generates light neutrino masses through the type-I see-saw mechanism~\cite{Senjanovic:2005seesaw,deAnda:2017yeb}. Leptogenesis can also be included, while its detailed realisation requires specifying the GUT-breaking and flavour sectors. We leave reheating and non-thermal wino DM for future work.

In summary, we  constructed a specific framework linking the CMB normalisation, high-scale SUSY breaking, Higgs mass generation and TeV-scale DM phenomenology. The inflation input fixes the characteristic high-energy scale, whereas the resulting low-energy spectrum gives testable predictions for direct detection and disappearing-track searches.


\section*{Acknowledgements}

The authors are grateful to Elisa Chisari, Andrei Egorov, Constantinos Pallis, Roman Pasechnik, Antonio Morais, Anca Preda and Tomislav Prokopec for discussions and correspondence.

 The revised \texttt{SusyHD} code used for RG running and related pictures is available in the ancillary files attached to arXiv:2607.07193.

 The work of DF was funded by the NWA ORC programme Emergence at All Scales. DF was also partially supported by the Foundation for Advancement of Theoretical Physics and Mathematics "BASIS".
 AB is  supported in part through the NExT Institute and STFC CG ST/X000583/1. AB acknowledges partial  support from Leverhulme Trust project MONDMag (RPG-2022-57). SVK was  partially supported by the Simons Foundation in the USA, and the World Premier International Research Center Initiative, MEXT, Japan. SVK is grateful to the International Institute of Physics in Natal, Brazil, for the kind hospitality extended to him during this investigation. 


\appendix
\renewcommand{\theequation}{\arabic{equation}}
\setcounter{equation}{0}
\renewcommand{\theHequation}{A.\arabic{equation}}

\section{Higgs mass uncertainties, and comparison of \texttt{SusyHD} and \texttt{HSSUSY}}
\label{app:higgs_uncertainties}

\texttt{SusyHD} provides three components of its intrinsic theory uncertainty: a low-energy SM component, an MSSM-to-SM matching component, and an EFT truncation component. Combining them linearly gives
\begin{equation}
\Delta_{\rm th}M_h^{\mathtt{SusyHD}}
=
 1.23~{\rm GeV}
\end{equation}
for both the bare and dynamically generated $\mu$ cases. The uncertainty is dominated by the low-energy SM component; the SUSY-matching and EFT truncation components are numerically negligible at this precision.

\texttt{HSSUSY} uses a more recent uncertainty prescription. Its low-energy component is estimated by varying the treatment of higher-order corrections in the extraction of the running top Yukawa coupling and by varying the renormalisation scale used in the Higgs pole mass calculation, while the high-scale matching uncertainty is evaluated separately. For those two cases, \texttt{HSSUSY} gives
\begin{equation}
\Delta_{\rm th}M_h^{\mathtt{HSSUSY}}
=
 0.526~{\rm GeV}
\quad {\rm and} \quad
 0.423~{\rm GeV}\,,
\end{equation}
respectively. The difference between the \texttt{SusyHD} and \texttt{HSSUSY} theory uncertainties reflects different perturbative uncertainty prescriptions, not a disagreement between their central predictions.

During the comparison, we found a bug in the precomputed fast interpolation table distributed with \texttt{SusyHD} v1.0.2. The table is parametrised by the strong gauge coupling $g_3$ at the top-mass scale, but its entries are numerically independent of this coordinate. In the original table generation, the RG boundary condition used the fixed central value $g_{3{\rm MT}}$ rather than the varying grid value $g_{3{\rm MT}}^{\rm var}$. As a result, a variation of $\alpha_s(M_Z)$ was propagated through the top Yukawa coupling, but its independent effect through $g_3$ was omitted, leading to an underestimated $\alpha_s$ parametric uncertainty.

We regenerated the interpolation table using the original \texttt{SusyHD} RG equations and the corrected boundary condition
\begin{equation}
g_3(0)=g_{3{\rm MT}}^{\rm var}.
\end{equation}
The resulting table has a non-zero dependence on $g_3$. We use this table in a local corrected version,  \texttt{SusyHD} v1.0.3 (30 June 2026). This is not an official upstream \texttt{SusyHD} release. The corrected source, the regenerated table, the regeneration script and a README file are provided as supplementary material. The public \texttt{SusyHD} interface and all input conventions are unchanged.

With the corrected table, the $\alpha_s$ uncertainties in \texttt{SusyHD} are
\begin{equation}
\Delta_{\alpha_s}M_h^{\rm \texttt{SusyHD}}
=
 0.641~{\rm GeV}
\quad {\rm and} \quad
 0.643~{\rm GeV}
\end{equation}
for the bare and dynamically generated $\mu$ cases, respectively. \texttt{HSSUSY} gives
\begin{equation}
\Delta_{\alpha_s}M_h^{\rm \texttt{HSSUSY}}
=
 0.644~{\rm GeV}
\quad {\rm and} \quad
 0.646~{\rm GeV}\,,
\end{equation}
respectively. The corrected \texttt{SusyHD} and \texttt{HSSUSY} calculations give almost identical $\alpha_s$ dependence. The correction leaves the central \texttt{SusyHD} masses and intrinsic theory uncertainties unchanged at the quoted precision because the central $g_3$ grid point was already evaluated correctly.

The Higgs mass calculation involves boundary conditions imposed at two widely separated scales. The measured top pole mass, the strong coupling and the remaining EW inputs are converted into running SM parameters at a low renormalisation scale. In particular, one has
\begin{equation}
M_t^{\rm pole}
\longrightarrow
m_t^{\overline{\rm MS}}(Q_{\rm EW})
\longrightarrow
y_t(Q_{\rm EW}) .
\end{equation}
The gauge and Yukawa couplings are evolved to the scale $m_0$ where the MSSM-to-SM matching condition fixes the boundary value of the SM Higgs quartic coupling as
\begin{equation}
\lambda(m_0)
=
\fracmm{1}{4}
\left[
g_2^2(m_0)
+
\fracmm{3}{5}g_1^2(m_0)
\right]
\cos^2 2\beta
+
\Delta\lambda(m_0) .
\end{equation}
For $\tan\beta=1$,
\begin{equation}
\cos 2\beta=0,
\end{equation}
so the tree-level contribution vanishes and
\begin{equation}
\lambda(m_0)
=
\Delta\lambda(m_0) .
\end{equation}
The quartic coupling and the other SM parameters are evolved to the EW scale, where the running parameters are converted into the physical Higgs pole mass.

The coupled low- and high-scale boundary conditions are solved iteratively. Iteration removes numerical inconsistencies in the boundary-value solution, but it does not remove the uncertainty due to missing perturbative orders. The relation between the measured top pole mass and the running top mass is known only as a truncated perturbative expansion,
\begin{equation}
M_t^{\rm pole}
=
m_t^{\overline{\rm MS}}(Q)
\left[
1
+
\Delta_t^{(1)}(Q)
+
\Delta_t^{(2)}(Q)
+
\Delta_t^{(3)}(Q)
+
\ldots
\right]
\end{equation}
Different prescriptions that are equivalent up to the known order can therefore give slightly different values of $y_t(Q_{\rm EW})$ after convergence.

The top Yukawa coupling strongly affects the RG evolution of the Higgs quartic coupling. Schematically one has,

\begin{equation}
16\pi^2\beta_\lambda
=
-12y_t^4
+
12\lambda y_t^2
+
\ldots .
\end{equation}

Consequently, a small higher-order difference in the extraction of $y_t(Q_{\rm EW})$  changes the evolution of $\lambda$ over the large interval
\begin{equation}
\ln\left(\fracmm{m_0}{M_t}\right)
\simeq 25 .
\end{equation}
The RG evolution does not introduce an independent source of uncertainty. Instead, the long running interval increases the sensitivity of the predicted low-energy Higgs mass to missing higher-order terms in the low-energy determination of $y_t$ and in the conversion between the running quartic coupling and the Higgs pole mass.

The intrinsic perturbative uncertainty should be distinguished from the parametric uncertainty induced by the experimental errors of the SM inputs. Varying
\begin{equation}
M_t=172.52\pm0.33~{\rm GeV}
\end{equation}
induces $\Delta_{M_t}M_h\simeq0.627~{\rm GeV}$ in both codes and for both cases.
 The $M_t$ uncertainty is kept separate in the present analysis because the dependence on $M_t$ is explicitly displayed  along the horizontal axis in Fig.~\ref{fig:higgs_mass_SusyHD_HSSUSY_comparison}. At fixed $M_t$, varying
\begin{equation}
\alpha_s(M_Z)=0.1180\pm0.00090
\end{equation}
gives the $\alpha_s$ uncertainties quoted above. The vertical bands in
Fig.~\ref{fig:higgs_mass_SusyHD_HSSUSY_comparison} are obtained by adding the
intrinsic theory and $\alpha_s$ uncertainties linearly.

\texttt{SusyHD} was introduced in 2015~\cite{PardoVega:2015eno}. Its latest publicly distributed release, v1.0.2, remains the 2015 implementation. Our local v1.0.3 correction changes only the fast interpolation table and restores the missing $g_3$ dependence; it does not update the perturbative content or the uncertainty prescription of the original code. The agreement of its central prediction with \texttt{HSSUSY} at the level of about $0.5~{\rm GeV}$ shows that the \texttt{SusyHD} central value remains a useful cross-check. However, its built-in uncertainty prescription is based on the perturbative information and uncertainty estimators implemented in 2015.

Since the release of \texttt{SusyHD}, substantial progress in high-scale SUSY Higgs mass calculations has been made. This includes improved EFT calculations, higher-order threshold corrections, improved treatment of the running SM parameters, and more systematic point-by-point estimates of missing higher-order effects~\cite{Athron:2016fuq,Athron:2017fvs,Bagnaschi:2017xid,Slavich:2020zjv}. The \texttt{SusyHD} result
\begin{equation}
\Delta_{\rm th}M_h^{\rm \texttt{SusyHD}}
\simeq
 1.23~{\rm GeV}
\end{equation}
should therefore be interpreted as a conservative legacy estimate associated with the 2015 \texttt{SusyHD} implementation. For the phenomenological interpretation, we adopt the more recent \texttt{HSSUSY} estimates,
\begin{equation}
\Delta_{\rm th}M_h^{\rm \texttt{HSSUSY}}
=
 0.526~{\rm GeV}
\quad {\rm and} \quad
 0.423~{\rm GeV},
\end{equation}
for the bare and dynamically generated $\mu$ scenarios, respectively. These estimates remain prescription-dependent and should not be interpreted as statistical confidence intervals~\cite{Slavich:2020zjv}.



\bibliography{Bibliography}{}
\bibliographystyle{utphys}

\end{document}